%% file: paper.tex
\newif\ifdraft\draftfalse

\documentclass[acmsmall,screen,nonacm]{acmart}
\setcopyright{none}

\usepackage{xcolor}
\usepackage{tcolorbox}
\usepackage{xspace}
\usepackage{caption}
\usepackage{relsize}
\usepackage[normalem]{ulem}
\newcommand{\explaincolor}[2]{{\raisebox{-.3ex}{\color{#1}\rule{1.2em}{.8em}}}~= #2}
\usepackage{enumitem}
\usepackage{tikz}
\usetikzlibrary{positioning, arrows.meta, fit, backgrounds, calc, shapes.geometric}

\definecolor{dkblue}{rgb}{0,0.1,0.7}
\definecolor{ltblue}{rgb}{0,0.1,0.7}
\definecolor{dkbluet}{rgb}{0,0.1,0.7}
\definecolor{ltbluet}{rgb}{0.8,0.8,1}
\definecolor{dkgreen}{rgb}{0,0.5,0}
\definecolor{ltgreen}{rgb}{0.8,1,0.8}
\definecolor{dkred}{rgb}{0.7,0,0}
\definecolor{ltred}{rgb}{1,0.8,0.8}
\definecolor{dkviolet}{rgb}{0.3,0,0.5}
\definecolor{ltviolet}{rgb}{1,0.8,1}
\definecolor{dkpurple}{rgb}{0.7,0,0.4}
\definecolor{ltpurple}{rgb}{1,0.8,1}
\definecolor{dkorange}{rgb}{0.8,0.4,0}
\definecolor{ltorange}{rgb}{1,0.8,0.6}
\definecolor{dkyellow}{rgb}{0.6,0.6,0}
\definecolor{ltyellow}{rgb}{1,1,0.6}
\definecolor{dkgray}{rgb}{0.4,0.4,0.4}
\definecolor{ltgray}{rgb}{0.9,0.9,0.9}
\definecolor{olive}{rgb}{0.4, 0.4, 0.0}
\definecolor{dkolive}{rgb}{0.4, 0.4, 0.0}
\definecolor{ltolive}{rgb}{0.9, 0.9, 0.0}
\definecolor{teal}{rgb}{0.0,0.5,0.5}
\definecolor{azure}{rgb}{0.0, 0.4, .8}
\definecolor{ltazure}{rgb}{0.8, 0.9, 1}
\definecolor{dkazure}{rgb}{0.0, 0.4, .8}

\definecolor{generator}{RGB}{75, 161, 241}
\definecolor{ltgenerator}{RGB}{75, 161, 241}
\definecolor{dkgenerator}{RGB}{75, 161, 241}

\definecolor{mutator}{RGB}{174, 62, 201}
\definecolor{ltmutator}{RGB}{174, 62, 201}
\definecolor{dkmutator}{RGB}{174, 62, 201}

\definecolor{shrinker}{RGB}{241, 172, 75}
\definecolor{ltshrinker}{RGB}{241, 172, 75}
\definecolor{dkshrinker}{RGB}{241, 172, 75}

\definecolor{checker}{RGB}{224, 49, 49}
\definecolor{ltchecker}{RGB}{224, 49, 49}
\definecolor{dkchecker}{RGB}{224, 49, 49}

\definecolor{printer}{RGB}{159, 168, 178}
\definecolor{ltprinter}{RGB}{159, 168, 178}
\definecolor{dkprinter}{RGB}{159, 168, 178}

\definecolor{feedback}{RGB}{76, 176, 94}
\definecolor{ltfeedback}{RGB}{76, 176, 94}
\definecolor{dkfeedback}{RGB}{76, 176, 94}

\definecolor{seedpool}{RGB}{248, 119, 119}
\definecolor{ltseedpool}{RGB}{248, 119, 119}
\definecolor{dkseedpool}{RGB}{248, 119, 119}

\definecolor{chart_black}{HTML}{000000}
\definecolor{chart_red}{HTML}{900D0D}
\definecolor{chart_orange}{HTML}{DC5F00}
\definecolor{chart_blue}{HTML}{243763}
\definecolor{chart_green}{HTML}{436E4F}
\definecolor{chart_purple}{HTML}{6D0E56}
\definecolor{chart_pink}{HTML}{D61C4E}

\definecolor{chart_heap_seed_pool}{HTML}{C62E2E}  
\definecolor{chart_filo_seed_pool}{HTML}{4C4B16}
\definecolor{chart_fifo_seed_pool}{HTML}{740938}
\definecolor{chart_resetting_seed_pool}{HTML}{3D0301}
\definecolor{chart_monotonic_seed_pool}{HTML}{000B58}
\definecolor{chart_static_seed_pool}{HTML}{CB6040}

\newcommand{\overlay}[2]{\raisebox{-0.1\height}{\tcbox[on line,boxsep=0pt, boxrule=1pt,left=2pt,right=2pt,top=1pt,bottom=1pt, colframe=dk#1, colback=lt#1]{\tt #2}}}
\newcommand{\overlayfull}[2]{\tcbox[on line,boxsep=0pt, boxrule=1pt,left=2pt,right=2pt,top=2pt,bottom=2pt, colframe=dk#1, colback=lt#1]{#2}}

\newcommand{\roverlay}[2]{\raisebox{0.3mm}{\overlay{#1}{#2}}}

\input{figures/interaction/tikz-loops}

\input{macros.tex}

\begin{document}

\title{Programmable Property-Based Testing}

\author{Alperen Keles}
\email{akeles@umd.edu}
\orcid{0009-0000-5734-3598}
\affiliation{%
  \institution{University of Maryland, College Park}
  \streetaddress{}
  \city{College Park}
  \state{Maryland}
  \country{USA}
  \postcode{20740}
}

\author{Justine Frank}
\email{jpfrank@umd.edu}
\orcid{0009-0007-7024-7331}
\affiliation{%
  \institution{University of Maryland, College Park}
  \streetaddress{}
  \city{College Park}
  \state{Maryland}
  \country{USA}
  \postcode{20740}
}

\author{Ceren Mert}
\email{cmert@umd.edu}
\orcid{0009-0002-9365-0661}
\affiliation{%
  \institution{University of Maryland, College Park}
  \streetaddress{}
  \city{College Park}
  \state{Maryland}
  \country{USA}
  \postcode{20740}
}

\author{Harrison Goldstein}
\email{hgoldste@buffalo.edu}
\orcid{0000-0001-9631-1169}
\affiliation{%
  \institution{University at Buffalo, SUNY}
  \streetaddress{}
  \city{Buffalo}
  \state{New York}
  \country{USA}
  \postcode{14260}
}

\author{Leonidas Lampropoulos}
\email{leonidas@umd.edu}
\orcid{0000-0003-0269-9815}
\affiliation{%
  \institution{University of Maryland, College Park}
  \streetaddress{}
  \city{College Park}
  \state{Maryland}
  \country{USA}
  \postcode{20740}
}


\begin{abstract}
  Property-based testing (PBT) is a popular technique for establishing
  confidence in software, where users write {\em
      properties}---i.e. executable specifications---that can
  be checked many times in a loop by a testing framework.
  In modern PBT frameworks, properties are usually written in {\em
      shallowly embedded} domain-specific languages, and their definition
  is tightly coupled to the way they are tested. Such frameworks often
  provide convenient configuration options to customize aspects of the
  testing process, but users are limited to precisely what library
  authors had the prescience to allow for when developing the
  framework; if they want more flexibility, they may need to write a new
  framework from scratch.

  We propose a new, deeper language for
  properties based on a mixed embedding that we call {\em deferred binding
      abstract syntax}, which reifies properties as a
  data structure and decouples them from the property runners that execute
  them. We implement this language in Rocq and Racket, leveraging the
  power of dependent and dynamic types, respectively. Finally, we
  showcase the flexibility of this new approach
  by implementing a variety of property runners in a shared framework, highlighting
  domain-specific testing improvements that can be unlocked by more
  programmable testing.
\end{abstract}




\maketitle

\newcommand{\qc}{\textsc{QuickChick}}
\newcommand{\fc}{\textsc{FuzzChick}}
\newcommand{\rc}{\textsc{RackCheck}}
\newcommand{\etna}{\textsc{Etna}}

\section{Introduction}\label{sec:intro}

In property-based testing~\cite{ClaessenH00}, users build confidence
in their code using {\em properties} that describe what it means for a
program to be correct, expressed in the form of universally quantified
executable predicates: e.g.
for all expressions, evaluating them with and without optimizations
yields the same result.
%
Property-based testing frameworks provide two main pieces that
facilitate testing: a {\em property language} and a {\em property
    runner}.
The property language is the API that allows users to
express these executable predicates, and it provides limited ways to
configure testing behavior (such as how to generate or pretty print
test inputs);
the property runner is the way a framework actually runs a property
(taking any configuration into account) to evaluate the system under
test.

Consider, for example, the optimization correctness property above
in Haskell's QuickCheck:

\begin{minipage}{0.53\textwidth}
  \begin{hask}
    prop_eval :: Property
    prop_eval =
    forAllShrinkShow $\textcolor{generator}{\tt gen}$ $\textcolor{shrinker}{\tt shrink}$ $\textcolor{printer}{\tt show}$
    $\textcolor{checker}{\tt(\backslash e \rightarrow eval \: e == eval \: (optimize \: e))}$
  \end{hask}
\end{minipage}
\begin{minipage}{0.45\textwidth}
  \begin{hask}
    $\textcolor{generator}{\tt gen}$      :: Gen Exp
    $\textcolor{shrinker}{\tt shrink}$ :: Exp -> [Exp]
    $\textcolor{printer}{\tt show}$  :: Exp -> String
    $\textcolor{checker}{\tt eval, \: optimize}$      :: Exp -> Exp
  \end{hask}
\end{minipage}

\noindent
Given some type of expressions \HC{Exp}, users provide (or automatically derive\cite{GeneratingGoodGenerators})
%
(1) a generator for expressions \textcolor{generator}{\tt gen}, that is, a function from some
random seed to a concrete \HC{Exp};
(2) a shrinking function \textcolor{shrinker}{\tt shrink}, a function from an expression to a
list of potentially smaller expressions for minimization purposes;
(3) a printing function \textcolor{printer}{\tt show}, in order to report counterexamples
to the user;
(4) and a predicate on expressions, which in this case is an
\textcolor{checker}{anonymous function}
that given an expression \HC{e}, evaluates it with and without optimizations,
and checks that the results are equal.
To create a \HC{Property} that QuickCheck can test, users can leverage
the \HC{forAllShrinkShow} combinator from QuickCheck's property language API
to put everything together, or let typeclasses take care of this final assembly.
%
%
%
%
QuickCheck, and most PBT frameworks, arm users with a variety of ways to
configure many aspects of testing: from
implementing their own hand-tuned generator that produces a better
distribution of inputs to collecting bespoke statistics to gauge the
distribution of inputs produced.

\begin{figure}[b]
  \centering
  \resizebox{0.95\linewidth}{!}{\input{figures/interaction/quickchick.tex}}
  \caption{An Abstract Representation of QuickCheck's Property Runner}
  \label{fig:quickchick-property-runner-intro}
\end{figure}

\begin{figure}[b]
  \centering
  \resizebox{0.95\linewidth}{!}{\input{figures/interaction/fuzzing.tex}}
  \caption{An Abstract Representation of FuzzChick's Property Runner}
  \label{fig:fuzzing-property-runner-intro}
\end{figure}

Unfortunately, there is one aspect of testing that is baked into these
frameworks and can fundamentally not be configured: the property
runner---the very way properties are tested.
The core of QuickCheck's property runner is surprisingly
simple, as we depict pictorially in
Fig.~\ref{fig:quickchick-property-runner-intro}. Expressions are
continuously generated and checked until a counterexample is found;
then the counterexample is repeatedly shrunk until the smallest expression that
invalidates the property is found. The
(locally) minimal counterexample is then presented to the user.
While some details of these two loops can be configured (e.g., the number of
tests to run, how long to spend on shrinking, etc.),
users cannot configure the core
functionality of the loop itself. As we will see, the structure of the testing loop is baked
into the way properties are represented internally as a shallow
embedding.

But there are good reasons that a user might want to change the
testing loop!  Recent advances offer compelling alternative
structures.  For example, the literature is rife with testing
approaches inspired by coverage-guided fuzzing, where rather than
generating new inputs from scratch at each iteration of the loop, the
runner instead keeps track of inputs that led execution down a novel
path, and mutates those in the hopes of uncovering yet more
interesting paths\cite{HypoFuzz,Crowbar,FuzzChick}. If we were to draw
a diagram for such a system, such as FuzzChick~\cite{FuzzChick}, we
might get something like the one that appears in
Fig.~\ref{fig:fuzzing-property-runner-intro}.

Despite the similarity of these two diagrams, the implementations of these
approaches are almost entirely distinct: no reuse of components for
the implementors of the frameworks, no reuse of properties for the
users.
As we will discuss in detail, the prevalent QuickCheck-based
design of the property language bakes the generation and minimization
loops into an opaque representation of properties using a shallow
embedding that only allows for a single predefined interpretation:
executing the loop of Fig.~\ref{fig:quickchick-property-runner-intro}.
%
%
In this work, we challenge this design decision, and introduce a new
way of representing properties that allows property runners to be
written at the ``user-level'', without needing to dive into or modify
library internals.

%


To that end, we turn to a deeper, but not entirely deep, embedding for
the representation of properties, designing a language that can be
reified as a data structure and then interpreted in different
ways~\cite{hudakBuildingDomainspecificEmbedded1996,PHOAS,goldsteinParsingRandomness2022,DeeperShallowEmbeddings}.
%
%
We introduce a novel style of mixed embeddings that is particularly
well-suited for representing properties for testing, which we call
  {\em deferred binding abstract syntax}: rather than binding a variable
once at the site of its universal quantification, we will instead bind
it at every one of its use sites.
This representation allows us to fully {\em decouple the specification from the
    runner}: the property itself expresses only the specification and the
runner can be programmed by the user to interpret the
property---allowing for maximum programmability.

We offer the following contributions:
\begin{itemize}[leftmargin=2em]
  \item We introduce a new style of mixed embeddings, which we call {\em deferred binding abstract syntax}, and use it to define a property language that allows for arbitrary re-interpretation of
        the way properties are executed (\S~\ref{sec:property-languages}).
  \item We implement our property language API in the QuickChick property-based testing
        framework for Rocq, leveraging dependent types to ensure that properties are well-formed
         while retaining a surface syntax close to existing property APIs (\S~\ref{sec:rocq}).
  \item We implement our property language API as part of a new PBT framework in Racket, leveraging dynamic typing and macros to hide the internal data structure, providing an
        identical user interface to existing libraries, while enabling
        flexibility through the deeper embedding (\S~\ref{sec:racket}).
  \item We thoroughly evaluate our approach in three ways: First, we
        demonstrate the flexibility of this new language by implementing a
        variety of complex runners from the recent literature---including ones
        with coverage-guided fuzzing and context-sensitive shrinking---all in
        user code.
        Then, we compare the performance of our implementation to existing
        frameworks, showing negligible overhead.
        Finally, we showcase how the flexibility of our approach allows
        experiments to be expressed as changes to reusable runners and runner
        components, by carrying out three experiments to fine tune testing aspects.
\end{itemize}

\noindent
We conclude with related (\S~\ref{sec:related}) and future
(\S~\ref{sec:future}) work.

\section{Background: Property Runners}\label{sec:runners}

We begin by motivating the need for a programmable property-based
testing framework by exploring a wide variety of property runners from
the literature, highlighting their similarities and differences. We
depict the runners using abstract representations (like the ones in
the introduction) of the dataflow between different \emph{components}
of the runners, such
as \textcolor{generator}{generators}, \textcolor{shrinker}{shrinkers},
or \textcolor{printer}{printers}.
Crucially, existing implementations of the runners we discuss are
distinct, spread across different libraries in
different languages. This section presents our attempt at putting them
within a single conceptual framework; later we will show how to
implement all of them in a single framework.

\subsection{Simple Generational Property Runner} \label{sec:loops:sgpr}

We have already discussed the quintessential property runner of
Fig.~\ref{fig:quickchick-property-runner-loops}, as proposed by
Quick\-Check~\cite{ClaessenH00}. Looking a bit more closely, the runner
consists of two stages: the first is a generate-and-check stage that
repeatedly generates random inputs and tests them against the stated
property until a maximum limit on the number of tests is reached, or
until an input falsifying the property---i.e, a bug---is found. If
such a bug is found, the second stage of the runner is a
shrink-and-check stage that repeatedly tries smaller inputs (produced
by a user-provided ``shrinking'' function) until a local minimum---an
input that can no longer be shrunk---is reached.

\begin{figure}[h]
  \centering
  \resizebox{0.95\linewidth}{!}{\input{figures/interaction/quickchick.tex}}
  \caption{The QuickCheck-style property runner from Fig.~\ref{fig:quickchick-property-runner-intro},
  repeated here for comparison with alternative runner structures.}
  \label{fig:quickchick-property-runner-loops}
\end{figure}

\noindent
In principle, such a runner could be implemented using two
straightforward modular loops that could be composed together in
sequence. However, QuickCheck's design has the shrinking behavior
built into the way properties are executed: when a test input is
generated, a tree of potentially smaller counterexamples is lazily
generated along the way, and is used for minimization. As such,
changing the shrinking behavior of QuickCheck, for example to
introduce integrated shrinking as we will discuss right below,
necessitates deep changes to the property representation and the
property runner, and as a consequence often a new framework.
%
%

\subsection{Integrated Shrinking Property Runner} \label{sec:loops:ispr}

Integrated shrinking emerges as a solution to a prevailing problem of
type-based shrinking: generators are usually tuned to only produce
inputs that satisfy implicit or explicit validity constraints, and
type-based shrinking most of the time will not take such constraints
into account, leading to minimized but invalid counterexamples. A
well-known instance of this issue manifested in the line of work on
Csmith and C-Reduce ~\cite{YangCER11, RegehrCCEEY12}, where C programs
were generated as to not exhibit undefined behavior, but
shrinking based on the C program grammar alone is likely to introduce undefined
behavior during minimization, producing uninteresting counterexamples unless the
shrinking process is guarded by external undefined-behavior detection tools.



Integrated shrinking solves this problem by removing the shrinker from
the equation altogether, instead reusing the generators themselves to
carry out the minimization. Generation can be thought of as a process
where random bytes are parsed into structured test
cases~\cite{goldsteinParsingRandomness2022}; in integrated shrinking,
rather than minimizing the structured output of generators, frameworks
minimize the randomness that goes into them. As a result, any validity
constraints that are built into a generator are preserved by
construction during shrinking.
%
%
%
This method of integrated shrinking, depicted in
Fig.~\ref{fig:rackcheck-property-runner}, can be found in many
frameworks such as Python's Hypothesis~\cite{HypothesisShrinking} and
Haskell's Hedgehog~\cite{Hedgehog} and Falsify~\cite{falsifyLeo2023}.

\begin{figure}[h!]
  \centering
  \resizebox{0.95\linewidth}{!}{\input{figures/interaction/rackcheck.tex}}
  \caption{An Abstract Representation of the Integrated Shrinking Property Runner}
  \label{fig:rackcheck-property-runner}
\end{figure}

Traditional and integrated shrinking offer an interesting
trade-off. Integrated shrinking removes the burden of writing
shrinkers (and especially ones that preserve any input invariants)
altogether, while QuickCheck's traditional approach can often lead to
substantially smaller inputs. In other words, different situations
call for different approaches. Unfortunately, ``baking in'' the way
inputs are minimized into the way properties are represented and
executed means that to take advantage of a different approach users
might have to switch frameworks, or even languages entirely! In
\S~\ref{sec:shrink-comp} we will show how the flexibility of our proposed
representation allows users to leverage whichever approach is more
appropriate, exploring this trade-off.

\subsection{Coverage-Guided (Fuzzing) Property Runner} \label{sec:loops:cgpr}

As we already discussed in the introduction, coverage-guided fuzzing,
popularized by AFL~\cite{aflplusplus}, is an effective random testing
technique which makes major, breaking changes to the generation loop,
as shown in Fig.~\ref{fig:fuzzing-property-runner}. In more detail,
the predicate that is being checked is instrumented, allows the runner
to keep track of the branches that were covered during execution; then,
the runner uses a genetic-style algorithm that attempts to maximize this coverage.
In most instantiations, the runner keeps track of a {\em seed pool}: a
corpus of inputs that led execution down interesting, previously
unseen paths. Inputs are then selected from this pool, mutated
randomly, until a bug (or, more commonly, simply a crash) is found.

\begin{figure}[h!]
  \centering
  \resizebox{0.95\linewidth}{!}{\input{figures/interaction/fuzzing.tex}}
  \caption{An Abstract Representation of the FuzzChick Property Runner}
  \label{fig:fuzzing-property-runner}
\end{figure}

The effectiveness of fuzzing has led to multiple attempts to
incorporate it into property-based testing: from early ones like
OCaml's Crowbar~\cite{Crowbar}, Rocq's FuzzChick~\cite{FuzzChick}, and
Java's JQF~\cite{Padye2019}, to more recent ones like
HypoFuzz~\cite{HypoFuzz}. Crucially, each such attempt was, by
necessity, either a completely new framework (Crowbar, JQF), or a
major redesign of an existing one (QuickChick for FuzzChick,
Hypothesis for HypoFuzz) that required significant changes to
internals in order to even offer an alternative to its existing
testing strategy.

Worse, the difficulty of incorporating such changes into existing
frameworks also makes it difficult to experiment with new ideas.  For
example, the search strategy of a fuzzing loop (how seeds are selected
for mutation) has seen significant research
interest~\cite{DBLP:journals/tse/BohmePR19, FairFuzz,
  DirectedGreybox}.  However, when adapting property-based testing to
incorporate fuzzing, FuzzChick reused AFL's choices~\cite{FuzzChick}
instead of experimenting with new options in the new higher-level
setting; in part because each option in such an
experimentation would effectively require a new framework. Situations
like this are precisely what a programmable and flexible runner
framework addresses; in \S~\ref{sec:seedpool}, we carry out an
extensive case study to explore what effect different search
strategies and seed pool experimentations have in testing.

\subsection{Custom-Feedback Guided (Targeted) Property Runner} \label{sec:loops:tpr}

Coverage information is not the only form of feedback that can be used to guide input generation.
The literature is rife with domain-specific metrics that can be used to produce interesting inputs:
in SlowFuzz~\cite{SlowFuzz} and PerfFuzz~\cite{PerfFuzz}, the goal of
the testing campaign is to discover algorithmic complexity
vulnerabilities that could lead to Denial of Service attacks in
systems, which is achieved by maximizing instruction counts and
execution path lengths;
in SQLancer~\cite{SQLancer}, the goal is to find bugs in SQL engines, achieved by maximizing the diversity
of the generated query plans~\cite{RiggerQPG23}.
It is also possible to consolidate such different feedback forms under
a single compositional design, as demonstrated by
FuzzFactory~\cite{FuzzFactory} and Target~\cite{TargetedPBT}, which
allows users to define and combine different feedback functions for
guiding the generation of inputs.
%
Fig.~\ref{fig:targeted-property-runner} depicts an abstract
representation of this last targeted property runner, where the
feedback is a separate function that is computed over the input
instead of a value obtained via instrumentation.

\begin{figure}[h!]
  \centering
  \resizebox{0.95\linewidth}{!}{\input{figures/interaction/targeted.tex}}
  \caption{An Abstract Representation of the Targeted Property Runner}
  \label{fig:targeted-property-runner}
\end{figure}

Once again, however, implementing such runners almost always meant implementing
an entirely new framework. Using our approach, implementing and experimenting
with different forms of feedback is available to users at the user-space of the library
without needing to dive into library internals.

\subsection{Combinatorial Property Runner} \label{sec:loops:cpr}

There are yet different forms of feedback that can be useful in
testing, that do not rely on dynamic feedback obtained during the
execution of a test, but rather on static features of generated
inputs. Combinatorial coverage offers one such
alternative~\cite{CombinatorialPBT, CombinatorialEnum} by performing
\emph{online generator thinning}, generating several inputs
at each iteration of the generate-and-check loop and only using the
input that covers the maximum amount of constructor interactions.
Such an approach, depicted in
Fig.~\ref{fig:combinatorial-property-runner}, is especially useful
when executing a test can be costly (e.g. when testing compilers),
and does not rely on mutation, and can also be encoded using
our approach.

\begin{figure}[t]
  \centering
  \resizebox{0.95\linewidth}{!}{\input{figures/interaction/combinatorial.tex}}
  \caption{An Abstract Representation of the Combinatorial Property Runner}
  \label{fig:combinatorial-property-runner}
\end{figure}

%

\subsection{Parallel Property Runner} \label{sec:loops:ppr}

Whereas the runners we explored so far focused on obtaining a better
distribution of inputs by guiding the generator towards "interesting"
parts of the search space, QuickerCheck~\cite{ParallelTesting} improved
upon the standard loop by exploiting parallelism: letting users run more
tests during the same time span.
In prior work, parallelization is made possible by building a "swarm"~\cite{swarm-testing} of
testing campaigns, each member focusing on a different subset of operations or capabilities
of the underlying system. QuickerCheck instead builds a parallel runner
where each thread cooperatively works within the same campaign.
Running independent QuickCheck campaigns naively can cause each worker to repeatedly
explore the same small-sized part of the input space. QuickerCheck avoids this by
sharing a growing size counter across workers.

In such a parallel runner, depicted in
Fig.~\ref{fig:parallel-property-runner}, different threads share a
common atomic size counter for ensuring parallel progress on each
thread while keeping the contention on shared resources to a
minimum. The paper presents near-linear speed-up for up to 6 threads
due to the ability of almost perfect responsibility sharing across
threads without contention. In \S~\ref{sec:parallel-experiment},
we discuss a re-implementation of that runner on top of a Racket
PBT library.

\begin{figure}[h!]
  \centering
  \resizebox{0.95\linewidth}{!}{\input{figures/interaction/parallel.tex}}
  \caption{An Abstract Representation of the QuickerCheck Property Runner}
  \label{fig:parallel-property-runner}
\end{figure}

\subsection{Discussion}

The property runners we discussed in this section have given rise to a
wide variety of frameworks across languages. Unfortunately, while
their pictorial representations appear similar, their implementations
are anything but: a tight coupling between property runners and
property DSLs means that implementation details often render it very
difficult, if not entirely impossible, to switch and experiment with
different approaches without major changes to the underlying code.
We discuss exactly how in the next section, before showcasing the
flexibility provided by our framework by implementing {\em all}
of the runners presented above in \S~\ref{sec:eval}.

%

\section{Property Languages}\label{sec:property-languages}

\newcommand*{\sprop}{\ensuremath{\mathit{prop}}}
\newcommand*{\sforall}{\ensuremath{\mathit{forall}}}
\newcommand*{\scheck}{\ensuremath{\mathit{check}}}
\newcommand*{\srun}{\ensuremath{\mathit{run}}}
\newcommand*{\tbool}{\ensuremath{\mathtt{Bool}}}
\newcommand*{\trand}{\ensuremath{\mathtt{Rand}}}
\newcommand*{\tfun}[2]{\ensuremath{#1 \rightarrow #2}}

Properties are simply \emph{universally quantified predicates}. We
want to define a layer of such properties on top of some standard host
language, equipped with at least booleans:
\[
  \begin{array}{rcl}
    \tau & := & \tbool ~|~ \tfun{\tau}{\tau}                     \\
    e    & := & x ~|~ \lambda x. e ~|~ e ~ e ~|~ T ~|~ F ~|~ ... \\
  \end{array}
\]
In practice, the host language needs to also support some kind of
randomness for generation, lists for shrinking, strings for pretty
printing, etc. For presentation purposes, let's begin by formally
describing the core boolean structure:
\[
  \begin{array}{rcl}
    p & := & \forall x : \tau. ~p ~|~ e \\
  \end{array}
\]
That is, properties are either universal quantifiers or an injection
of a host predicate.

The core question we ask in this paper is: \emph{how should we
  represent this language?} We'll start by reviewing existing embedding
solutions from the literature, demonstrating why they are not flexible
enough to allow for users to customize or specify their runners,
before proposing our solution.
%

\subsection{Background Terminology: Deep and Shallow Embeddings}

An \emph{embedding} involves two languages: an \emph{object language}
that is being implemented, and a \emph{host language} that the object
language is being implemented in. In a deep embedding, the terms of
the object language are represented as inductive data in the host
language. These representations allow for arbitrarily manipulating and
inspecting the terms of the object language via the usual mechanism of
pattern matching.  By contrast, shallow embeddings directly expand the
constructs of the object language in terms of constructs of the host
language, which allows for straightforward implementation but no
user-level term manipulation~\cite{boulton1992experience}. Naturally,
multiple hybrid alternatives have been explored, trading off between
the simplicity of implementation of shallow embeddings and the
flexibility of deep ones~\cite{DeeperShallowEmbeddings,PHOAS,Tlon}.
%
%
%

\subsection{Status Quo: A Shallow Property Language} \label{sec:shallow}

Defining a property language requires a construct for defining its inputs via universal quantification ($\forall$),
a construct for defining the boolean predicate that is validated through the property ($ check $), and finally
a mechanism for running the property. In PBT libraries using a shallow embedding following QuickCheck's design,
the mechanism is using host language lambdas as binders for the universal quantification.

\[
  \begin{array}{lcl}
    \sforall & : & \forall \tau. ~ (\tfun{\tau}{\sprop}) \rightarrow \sprop \\
    \scheck  & : & \tbool \rightarrow \sprop                                \\
    \srun    & : & \sprop \rightarrow \mathtt{IO}~()
  \end{array}
\]
Concretely, the type of properties is some type $\sprop$ in the host
language (often restricted via typeclass-like mechanism), $\scheck$
injects a host boolean into this type, and $\sforall$ takes a
host-level function and turns it into a property.  Finally, frameworks
also provide a $\srun$ function to test properties constructed using
this API.

The main advantage of this approach is that it is very convenient for
users to write properties (using host-level binders), and extremely
easy to test them (simply invoking the framework-provided $\srun$).
Implementing such a framework as a developer is also
straightforward--- a minimal implementation of this shallow API in
Haskell using splittable pseudorandomness~\cite{ClaessenP13} is shown
in Fig.~\ref{fig:core-qc-impl}: the type of generators for some
type \texttt{a} is simply a wrapper around a function from some source
of randomness to \texttt{a}; properties are generators
for \texttt{Result}s (a datatype that encodes at least whether testing
succeeded); \texttt{forall} takes a generator and a predicate and
binds them together taking care of the underlying randomness; and
the \texttt{run} loop simply generates and checks a result up to some
predefined limit on the number of tests, taking care again of the
underlying randomness, and reporting success or failure in the end.

\begin{figure}[t]
  \begin{hask}
    newtype Gen a = Gen (StdGen -> a)

    type Property = Gen Result
    data Result = OK | Failed

    forall :: Gen a -> (a -> Property) -> Property
    forall (Gen g) f = Gen (\r -> let (r1,r2) = split r
    Gen h = f (g r1)
    in h r2

    run :: Property -> IO ()
    run (Gen g) = do newStdGen >>= loop numTests
    where loop 0 _ = ... -- report success
    loop n r = let (r1, r2) = split r in case g r1 of
    OK     -> loop (n-1) r2
    Failed -> ... -- report counterexample
  \end{hask}
  \caption{An implementation of QuickCheck's core functionality.}
  \label{fig:core-qc-impl}
\end{figure}

In practice, implementations of such frameworks are more involved to
account for better reporting, minimizing counterexamples,
etc. However, a key characteristic of this approach is that
the \HC{Property} type is opaque to users: it's simply a wrapper
around a function. As a result, users cannot inspect its structure or
customize how properties are tested without modifying the definition
of \HC{Property} and the internals of the framework.


\subsection{Hypothetical: A Deep Property Language}

The polar opposite of the current shallow embedded paradigm would be
to use a deep embedding: define an inductive representation of both
the host language and the language of properties and require users to
program properties in that. That would mean that users of the testing
framework would have to learn {\em yet another} language in which to
write their specifications, foregoing all of the convenience that
host-language binders provide. Unsurprisingly, no frameworks have
taken this route: it's hard enough to convince users to write
properties and specifications without additional burdens to
adoption~\cite{goldsteinPropertyBasedTestingPractice2024}.


\subsection{Proposal: A Mixed Property Language}\label{sec:dbas}

With pure shallow and deep embeddings being unsatisfactory, we turn to
hybrid approaches to attempt to regain some amount of inspection and
manipulation capabilities while keeping the property-writing process
as ergonomic as possible. Researchers have explored a wide range of
  {\em mixed embeddings} that reify parts of the language being
considered. For example, we could attempt a HOAS-style approach, where
we define an inductive type of properties, with a constructor
corresponding to $\sforall$ and $\scheck$, both indexed by the
arguments of their shallow counterparts:

\newcommand*{\Ind}{\ensuremath{\mathit{Inductive}}}
\newcommand*{\mforall}{\ensuremath{\mathit{Forall}}}
\newcommand*{\mcheck}{\ensuremath{\mathit{Check}}}
\newcommand*{\mprop}{\ensuremath{\mathit{Prop}}}
\[
  \begin{array}{l}
    \Ind ~ \mprop :=                                            \\
    |~ \mforall : (\tau \rightarrow  \mprop) \rightarrow \mprop \\
    |~ \mcheck  : \tbool \rightarrow \mprop                     \\
  \end{array}
\]
\smallskip

Such a representation still allows us to encode the universal
quantifiers using host language binders, but also makes some headway
into the issue at hand---given a property we can now pattern match
against it! However, there is still a problem: to access the ``rest''
of the property that follows a universal quantification, we need an
element of the type being quantified over; since such elements are
going to be randomly generated, we can't actually recurse down the
structure of the property without restricting ourselves to something
along the lines of QuickCheck's \HC{Gen}erator monad.

The key issue is that we're hiding the definition of the ``rest'' of
the property under a host-language binder: the $Forall$ constructor
takes a function as an argument, which we cannot pattern match on to
go deeper. Could we use standard host language constructs to define
the predicates at the leaves of the property (the $\mcheck$s), while
retaining the ability to pattern match on the property structure?

\paragraph*{Proposal: Deferred Binding Abstract Syntax}

Our solution is what we call {\em deferred binding abstract syntax}
(DBAS): rather than binding a variable once at the site of its universal
quantification, we will instead bind it at every one of its use sites.

\newcommand*{\menv}{\ensuremath{\mathit{env}}}
\newcommand*{\mType}{\ensuremath{\mathit{Type}}}
\[
  \begin{array}{l}
    \Ind ~ \mprop ~ (\menv : [\mType]) :=                                                      \\
    | ~ \mforall : \forall \tau. ~ \mprop ~ (\tau :: \menv) \rightarrow \mprop ~ \menv         \\
    | ~ \mcheck  : (\llbracket \menv \rrbracket \rightarrow \tbool) \rightarrow \mprop ~ \menv \\
  \end{array}
\]

\noindent
We still define an inductive type of properties, with one constructor
for each combinator in our API. We also index our type of properties
by an environment: a list of types that have already been quantified.
The key change is that we move the host level binder from the binding
site ($\mforall$) to its use site (the $\mcheck$). That is, the
argument to $\mforall$ is no longer a function, but simply a property
with an extended environment; on the other hand, the argument to
$\mcheck$ is no longer a simple boolean, but a function that binds
  {\em everything} in the environment. In the code above, we denote that
as $\llbracket \menv \rrbracket$---we will see how it can be
implemented in a statically typed or dynamically typed setting
respectively in later sections.

At first glance, this is a counter-intuitive trade-off: every time you
want to use a variable, you have to bind {\em everything} that was
quantified before that point. That would only make sense in a
scenario where there are a lot of quantifications and few variable
uses---which is precisely the case for the language of properties!
Compared to the previous language representations, this DBAS-based one
allows us to access the structure of the ``rest'' of the predicate
without having access to a concrete value, which is necessary when
such values are to be randomly generated. Finally, compared to the
fully shallow representation, the added inductive structure and typing
information pose some burden to the user experience, but once again,
we'll address these issues in host-language-specific settings.

\paragraph{Adding Annotations}

Generalizing the property language above to include annotations for
generation, shrinking, or printing of individual elements is
straightforward. We simply include an optional extensible list of
annotations at the $\mforall$ constructor:
\[
  \begin{array}{l}
    \Ind ~ \mprop ~ (\menv : [\mType]) :=                                                                \\
    | ~ \mforall : \forall \tau. ~as \rightarrow ~ \mprop ~ (\tau :: \menv) \rightarrow \mprop ~ \menv   \\
    | ~ \mcheck  : (\llbracket \menv \rrbracket \rightarrow \tbool) \rightarrow \mprop ~ \menv           \\
    \\
    as := \emptyset ~|~ (\mathit{k}, \forall \tau. ~ \llbracket \menv \rrbracket \rightarrow \tau) :: as \\
    k ~ := \mathit{gen} ~|~ \mathit{shr} ~|~ \ldots                                                      \\
  \end{array}
\]
At a high level, we can annotate each $\mforall$ constructor with a
(possibly empty) sequence of (host-level) functions that quantify over
the context so far (as in $\mcheck$) and return annotation-specific
terms (e.g. a generator or a shrinker).  We demonstrate the exact
implementation of such annotations in the following two
host-language-specific sections.

\section{A Dependently Typed Property Language}
\label{sec:rocq}

Implementing the language of universally quantified properties using
deferred binding abstract syntax is straightforward on top of a
dependently typed language, but achieving good ergonomics can be a
challenge. In this section we will focus on implementing such a
language on top of the QuickChick~\cite{LeoThesis2018} framework for
property-based testing in Rocq.

To that end, we explicitly encode contexts in our properties,
capturing every input that {\em will have been generated} by that
point. We will use a standard inductive definition of contexts,
using \CC{\emp} to denote the empty context and \CC{\.} to extend a
context by a type.
Given a context, we can calculate the type corresponding to it: the
type of tuples containing all of its types in order, with \CC{unit} as
the base case:

\noindent \begin{minipage}{0.50\textwidth}
  \begin{rocqcode}
    Inductive Ctx :=
    | \emp : Ctx
    | \. : Type -> Ctx -> Ctx.
  \end{rocqcode}

  \vspace{2mm}

  We will write \CC{[[C]]} as a shorthand for \CC{interp C}.
\end{minipage}
\noindent \begin{minipage}{0.50\textwidth}
  \begin{rocqcode}
    Fixpoint interp (C : Ctx) : Type :=
    match C with
    | \emp => unit
    | T \. C => T * interp C
    end.
  \end{rocqcode}
\end{minipage}

Now we can define a deeper version of the property language using
DBAS:
\begin{rocqcode}
  Inductive Prop : Ctx -> Type :=
  | FORALL : forall {A: Type} {C: Ctx} (name: string)
  (generator : [[C]] -> G A)             (mutator  : [[C]] -> A -> G A)
  (shrinker   : [[C]] -> A -> list A) (printer   : [[C]] -> A -> string),
  Prop (A · C) -> Prop C
  | IMPLIES : forall C
  (prop : [[C]] -> bool),
  Prop C -> Prop C
  | CHECK : forall C,
  ([[C]] -> bool) -> Prop C.
\end{rocqcode}
Just like the shallow approach of QuickCheck, this representation
allows us to express, in the host language, type-based generators,
mutators, shrinkers, and printers for each quantifier in a
property. Just like the shallow approach, we can use typeclasses to
automate much of the burden of specifying the property (as we will see
below). Crucially, however, unlike the shallow approach we can pattern
match on this definition and construct a wide range of methods for
interpreting such properties without needing to modify the code of the
underlying property-based framework at all.

For example, the standard ``generate-and-run'' loop of
Figure~\ref{fig:quickchick-property-runner-intro}, which amounts to
interpreting such a property in the original shallow embedding, can be
straightforwardly encoded as follows, first defining a simple type of
  {\em results} that holds information such as the \CC{inputs} that were
generated (calculated recursively from an input property \CC{prop}):
\begin{rocqcode}
  Inductive RunResult {C: Ctx} (prop : Prop C) :=
  | Normal $~$: [[inputs prop]] -> bool -> RunResult prop
  | Discard : [[inputs prop]] -> RunResult prop.
\end{rocqcode}

\noindent
And then the runner as a straightforward fixpoint as shown in Fig.~\ref{fig:dbas-rocq-runner}:
\footnote{Branches follow the standard
  \href{http://adam.chlipala.net/cpdt/html/MoreDep.html}{convoy pattern} to
  enable type inference in dependent pattern matching, and we hide those
  in $\color{lightgray}{\mathit{gray}}$---they are unfortunate artifacts
  of Rocq's support for dependently typed programming.}

\begin{figure}[h]
  \begin{rocqcode}
    Fixpoint genAndRun (C : Ctx) (prop : Prop C) : [[ C]] -> G (RunResult prop) :=
    match prop with
    | FORALL A C name gen mut shr pri prop =>
    $\color{lightgray}{\mathtt{(fun~A'~C'~name'~gen'~mut'~shr'~pri'~prop'~=>}}$
    fun env =>
    a <- gen$\color{lightgray}{'}$ env;;
    res <- genAndRun (A$\color{lightgray}{'}$ · C$\color{lightgray}{'}$) prop$\color{lightgray}{'}$ (a, env);;
    match res with
    | Normal env truth => $\color{lightgray}{\mathtt{(fun~env'~truth'~=>}}$
    ret (Normal (Some a, env$\color{lightgray}{'}$) truth$\color{lightgray}{'}$)) $\color{lightgray}{\mathtt{env~truth}}$
    | Discard env => $\color{lightgray}{\mathtt{(fun~env'~=>}}$
    ret (Discard (Some a, env$\color{lightgray}{'}$))) $\color{lightgray}{\mathtt{env}}$
    end) $\color{lightgray}{\mathtt{A~C~name~gen~mut~shr~pri~prop}}$
    | CHECK C prop => $\color{lightgray}{\mathtt{(fun~C'~prop'~=>}}$
    fun env =>
    ret (Normal tt (prop$\color{lightgray}{'}$ env))) $\color{lightgray}{\mathtt{C~prop}}$
    | IMPLIES C pre prop => $\color{lightgray}{\mathtt{(fun~C'~pre'~prop'~=>}}$
    fun env =>
    if pre$\color{lightgray}{'}$ env then
    res <- genAndRun C$\color{lightgray}{'}$ prop$\color{lightgray}{'}$ env;;
    match res with
    | Normal env truth => $\color{lightgray}{\mathtt{(fun~env'~ truth'~=>}}$
    ret (Normal env$\color{lightgray}{'}$ truth$\color{lightgray}{'}$)) $\color{lightgray}{\mathtt{env~truth}}$
    | Discard env => $\color{lightgray}{\mathtt{(fun~env'~=>}}$
    ret (Discard env$\color{lightgray}{'}$) $\color{lightgray}{\mathtt{env}}$
    end)
    else ret (Discard (nones prop$\color{lightgray}{'}$))) $\color{lightgray}{\mathtt{C~pre~prop}}$
    end.
  \end{rocqcode}
  \caption{Implementation of a DBAS-powered generate-and-run loop in Rocq}
  \label{fig:dbas-rocq-runner}
  \vspace{-2mm}
\end{figure}
%
%
%
The flexibility to define such a loop at the hands of users allows for
encoding all kinds of interpreters for properties, including pure
generators, runners, shrinkers, fuzzers, with fully programmable
execution, printing, and benchmarking options. We'll further
demonstrate this flexibility by implementing a series of runners from
the literature, all on top of this abstract property language.

\vspace{-2mm}

\paragraph*{Usable Defaults with Dependently Typed Programming}

A standard disadvantage of deep embeddings compared to shallow ones,
is that they are generally less convenient to work with. Encoding
everything in the host language, as in a shallow embedding, allows
users to simply reuse a large part of host language
infrastructure. The property language described above enables much of
that using dependent types. Still, it is desirable to provide as
seamless an experience for new users as possible, leveraging the same
familiar typeclass-based interfaces of the shallow setting.

For concreteness, without any effort to provide such an experience,
users would have to write the following to encode the optimization
correctness property of the introduction:
\begin{rocqcode}
  Definition prop_eval_bad :=
  FORALL (fun tt => gen) (fun tt => mut)
  (fun tt => shrink) (fun tt => pretty) (
  @CHECK (Expr \. \emp) (fun '(e, _) => eval e == eval (optimize e))).
\end{rocqcode}
That is, users would have to write a lot of annotations to achieve the
same result, both at the type level (\CC{Expr \. \emp}) and to
annotate individual \CC{FORALL}s with the various generators,
shrinkers, and printers.

However, we are not restricted to providing the core property
definition as the final user-level interface. Instead, we develop a
simple surface-level language that allows users to write simplified
properties, using typeclasses to fill in the remaining
information. For example, the same property can be defined in our
framework as in the much more straightforward snippet that follows:
\begin{rocqcode}
  Definition prop_eval :=
  ForAll $e$ :- Expr,
  Check (fun '$e$ => eval e == eval (optimize e)).
\end{rocqcode}

In addition, users can override particular aspects of the property
with lightweight annotations.  For example, specifying a particular generator to be used,
such as \CC{gen} can be done as follows:
\begin{rocqcode}
  Definition prop_eval :=
  ForAll $e$ :- Expr gen:gen,
  Check (fun '$e$ => eval e == eval (optimize e)).
\end{rocqcode}

Most of this surface language is achieved using Rocq's powerful
notation mechanism, including its support for recursive notations.
The final piece of the puzzle to simplify \CC{Check} definitions
relies on typeclasses. In particular, we associate each predicate with
its corresponding context and a proof of that correspondence, in a
class we name \CC{Untuple}:
\begin{rocqcode}
  Class Untuple (A : Type) :=
  { untuple : Ctx
  ; untuple_correct : [[untuple]] = A }.
\end{rocqcode}
We then provide instances for the empty context and the bind:
\begin{rocqcode}
  Instance Untuple_empty : Untuple unit :=
  { untuple := \emp
  ; untuple_correct := eq_refl }.
  Instance Untuple_pair {A B} `{Untuple B} : Untuple (A * B) :=
  { untuple := A \. @untuple B _
  ; untuple_correct := ... }.
\end{rocqcode}
Before finally providing a convenient user-level wrapper for
property conclusions, which we used above:
\begin{rocqcode}
  Definition Check {A} `{Untuple A} (p : A -> bool) : Prop (@untuple A _).
  refine (Check (@untuple A _) _).
  rewrite untuple_correct.
  exact p.
  Defined.
\end{rocqcode}

Finally, we can also leverage the extensive metaprogramming facilities
of QuickChick to construct a property definition directly from a
predicate:
\begin{rocqcode}
  Definition eval_correct (e : Expr) := eval e == eval (optimize e).
  Derive Property eval_correct.
  (* ==> eval_correct_prop is defined. *)
\end{rocqcode}
This command constructs the deeper property above using
the Rocq predicate itself.

\section{A Dynamically Typed Property Language}
\label{sec:racket}

Representing the property language with deferred binding abstract
syntax is not restricted to a dependently typed setting; in this
section, we show how to implement it in a dynamically typed language
like Racket.  In such a setting, we no longer have static guarantees
about the types of the variables in the context; instead, we fall back
on dynamic errors.  However, we are also not encumbered by the type
system, as we can freely invoke functions on arguments of (statically)
unknown types, which we will fully take advantage of to recover most
of the convenience of a shallow representation.


The first step is to directly translate the datatype into a series
of \RC{struct}s:
\begin{racketcode}
  (struct Forall (var augments body))
  (struct Implies (prop body))
  (struct Check (prop))
\end{racketcode}
In Racket, we cannot rely on typeclasses to automatically discover
generators or shrinkers for property-defined variables.
Instead, we add a dictionary to the \RC{Forall}s that allows us to
attach extra information onto each variable that we call {\em augments}.
Concretely, this dictionary maps augment names to functions that take the
current environment of previously-generated variables and produce the
augment value. We write augment names as Racket keywords: identifiers
prefixed with \RC{#:} that serve as named dictionary keys.
These augments are fully generic in that they can store any
values, though our implementation defines specific uses for three.
\begin{itemize}
  \item \RC{#:contract} attaches an invariant contract: a predicate checked
          dynamically against each generated value bound to the variable.
  \item \RC{#:gen} attaches a generator for the variable.
        In order to make the usage of generators from other frameworks e.g. \rc{}
        easier, we intentionally treat the generator value as opaque.
        Instead, property interpretations that use the generators take a
        user-provided sampling function that is applied to the generator.
  \item \RC{#:shrink} attaches a function used for shrinking counterexamples.
\end{itemize}

However, handling the struct-based definitions directly involves a lot
of explicit plumbing that we would rather not need to write. Consider
once again the optimization correctness property:

\begin{racketcode}
  (define eval-opt
  (Forall 'e (hash '#:contract (lambda (env) expr?) '#:gen (lambda (env) gen-expr))
  (Check (lambda (env)
  (let ([e (dict-ref env 'e)])
  (equal? (eval e) (eval (optimize e))))))))
\end{racketcode}

There are two main ergonomic issues: the repeated nesting and the
explicit environment passing and lookup. We can use Racket's extensive
macro capabilities to create a DSL for writing these deeper
properties. To that end, we flatten the nested structure by using the
fact that properties are isomorphic to a list of \RC{Forall}
and \RC{Implies} terminated by a single \RC{Check}. Then, we use
Racket's variable transformer macros to define and refer to generator
variables as Racket identifiers and insert the dictionary passing
plumbing for us.


These features allow us to translate the above roundtrip property
into one much closer to how shallow embedding properties are written.

\begin{racketcode}
  (define eval-opt
  (property
  (forall e #:contract expr? #:gen gen-expr)
  (equal? (eval e) (eval (optimize e)))))
\end{racketcode}


\paragraph*{Property Runners}
Developing property runners in the Racket setting shares much of the
structure of the Rocq version, with some extra logic to attach the
contract to the generated value if present.

\begin{figure}[h]
  \begin{racketcode}
    (define (gen-and-run p sample . args)
    (let loop ([p p] [env (hash)])
    (match p
      [(Forall var augments body)
        ; Ensure the variable has a generator augment
        (unless (dict-has-key? augments '#:gen) (error 'no-generator))
        ; Generate a value using the sample function
        (define val (apply sample ((dict-ref augments '#:gen) env) args))
        ; Check the contract if present
        (when (dict-has-key? augments '#:contract)
        (invariant-assertion ((dict-ref augments '#:contract) env) val))
        ; Recur
        (loop body (dict-set env var val))]
      [(Implies prop body)
        (if (prop env)                 ; Check precondition
        (loop body env)            ; If it passes, recur
        (values 'discard env))]    ; If it fails, discard
      [(Check prop)
        (if (prop env)                 ; Check result
        (values 'pass env)         ; Success
        (values 'fail env))])))    ; Failure
  \end{racketcode}
  \caption{Implementation of a DBAS-powered generate-and-run loop in Racket}
\end{figure}

Encoding properties using deferred binding abstract syntax in Racket
gives users the same variety in choices of property runners loops that
the Rocq version does, as well as the same programmability for expert
users. We utilize Racket's contracts to optionally allow users to
enforce typed boundaries on generated variables. Racket's powerful
macros enable us to write properties in a style that requires little
syntactic overhead compared to shallow embeddings without sacrificing
any of the programmability provided by deeper embeddings. In the next
section, we show that the extra programmability does not come at a
performance cost, enables the writing of property runners that execute
many runs in parallel, and shrinking loops that find significantly
smaller counterexamples.

\section{Evaluation}
\label{sec:eval}

In this section we evaluate our approach and implementation:

\begin{enumerate}

  \item First, we demonstrate the expressive power of DBAS, showing that
        it allows for building new property runners without any dependence or
        modifications to the library internals: we implemented all of the
        property runners we presented in \S~\ref{sec:runners}. Here, we
        discuss two such runners and their implementations in detail, the
        standard QuickCheck-inspired property runner (\S~\ref{sec:loops:sgpr})
        and the mutation-based coverage-guided runner
        (\S~\ref{sec:loops:cgpr}). We provide the rest of the implementations
        as supplementary material in Appendix~\ref{sec:appendix}.

  \item Second, we evaluate the performance overhead of DBAS-based
        implementations, showing that it is comparable to their shallowly
        embedded counterparts. In \S~\ref{sec:shallow_vs_deep} we compare the
        performance of the DBAS-based property runner as implemented in Rocq
        and Racket to the existing property runners of \qc{} and \rc{}. We
        found that using the DBAS-based embedding has {\em no observable
            performance overhead}.

  \item Finally, we showcase how the added expressivity allows
        these experiments to be expressed as changes to reusable runners and runner
        components, rather than as changes to the property language or framework internals.
        We carry out three new
        experiments: In \S~\ref{sec:seedpool} we explore how different design
        choices with respect to the representation and sampling of the seed
        pool affect testing performance, improving \fc{} in the process;
        in \S~\ref{sec:shrink-comp}, we compare the default integrated
        shrinking capabilities of \rc{} with a simple external shrinker we
        implemented for a DBAS-style Racket library;
        in \S~\ref{sec:parallel-experiment}, we implement and benchmark a
        parallel property-runner inspired by a recent work on parallelizing
        QuickCheck~\cite{ParallelTesting}.

\end{enumerate}

\subsection{Property Runners with DBAS} \label{sec:property-runners}

We demonstrate the expressiveness of DBAS by implementing
all property runners from \S~\ref{sec:runners}. In
the last two sections we saw how to implement a
simple \texttt{genAndRun} loop in both Rocq and Racket. Leveraging the
ability to pattern match on properties and recurse down their
structure, we can similarly develop building blocks that will
help implement all property runners. In particular, for this section,
we'll use a $\overlayfull{generator}{generator}$, which produces
random inputs for a property;
a $\overlayfull{mutator}{mutator}$, which given inputs to a property mutates
some or all of them;
a $\overlayfull{checker}{runner}$, which given inputs to a property executes it;
a $\overlayfull{shrinker}{shrinker}$, which given a way to minimize a property's inputs
and a counterexample finds a locally minimal counterexample using best-first search;
and a $\overlayfull{printer}{printer}$, for reporting counterexamples.
The underlying pattern matching ability is still useful in the
presence of these abstractions. For example, we will also write
domain-specific versions of these functions for some of our
evaluation, such as
an \roverlay{feedback}{\roverlay{checker}{instrumentedRunner}} that
runs a property collecting instrumentation feedback in the process.

\subsubsection{QuickCheck-Style Property Runner}

Using the components outlined above we can implement any variation of
the standard generational runner, with code that closely follows its
depiction. Fig.~\ref{fig:simple-generational-testing-loop-rocq} shows
the implementation of the basic generate-and-test then shrink-and-test
property runner.
Fig.~\ref{fig:simple-generational-testing-loop-racket} shows the same
runner in Racket, with minor syntactical adjustments.

\noindent \begin{minipage}{0.57\textwidth}
  \begin{rocqcodesmaller}
    Definition runLoop (fuel : nat) (cprop : Prop \emp) :=
    let fix runLoop' (fuel : nat) (cprop : Prop \emp)
    (passed : nat) (discards: nat) : G Result :=
    match fuel with
    | O => ret (mkResult discards false passed [])
    | S fuel' =>
    $\overlayfull{generator}{input <- gen cprop (log2 (passed + discards));;}$
    $\overlayfull{checker}{res <- run cprop input ;;}$
    match res with
    | Normal seed false => (* Fails *)
    $\overlayfull{shrinker}{let shrunk := shrinker 10 cprop seed in}$
    $\overlayfull{printer}{let printed := print cprop 0 shrunk in}$
    ret (mkResult discards true (passed + 1) printed)
    | Normal _ true => (* Passes *)
    runLoop' fuel' cprop (passed + 1) discards
    | Discard _ _ =>   (* Discard *)
    runLoop' fuel' cprop passed (discards + 1)
    end
    end in
    runLoop' fuel cprop 0 0.
  \end{rocqcodesmaller}
  \captionsetup{justification=raggedleft}
  \captionof{figure}{Simple Generational Property Runner in Rocq}
  \label{fig:simple-generational-testing-loop-rocq}
\end{minipage}
\begin{minipage}{0.43\textwidth}

  \indent
  Each component of the property runner in Fig.~\ref{fig:quickchick-property-runner-intro} has a clear
  correspondence to the corresponding overlaid sections in Fig.~\ref{fig:simple-generational-testing-loop-rocq}.
  The functions \roverlay{generator}{gen}, \roverlay{checker}{run},
  \roverlay{shrinker}{shrinker}, and \roverlay{printer}{print} are the building
  blocks of user level property-runners, but can also be written by
  users themselves in a straightforward manner as shown in
  Sections~\ref{sec:rocq} and~\ref{sec:racket}. We use them here to
  present runners at a higher level of abstraction as enabled by DBAS,
  focusing on how these components interact with each other in order to
  showcase users how to implement new runners corresponding to their needs.

  The runner itself consists of two tight loops, the first one
  running \roverlay{generator}{gen}, \roverlay{checker}{run}, \roverlay{generator}{gen}, \roverlay{checker}{run}...
  until a counterexample is found, or until a predefined limit of tests
  has been reached. The second loop
  runs \roverlay{shrinker}{shrink}, \roverlay{checker}{run}, \roverlay{shrinker}{shrink}, \roverlay{checker}{run} \dots
  until it is not able to minimize the input further, reporting the
  smallest input within the shrinking process.

\end{minipage}

\vspace{-5mm}

\begin{figure}[h]
  \begin{racketcodesmaller}
    (define (run-loop tests p)
    (let loop ([n 0] [passed 0] [discards 0])
    (if (= n tests) (result #f passed discards #f)
    (let ($\overlayfull{generator}{[env (generate p run-rackcheck-gen (floor (log n 2)))]}$)
    (case $\overlayfull{checker}{(check-property p env)}$
    [(fail) (result #t passed discards $\overlayfull{shrinker}{(shrink-eager p env)}$)]
      [(pass) (loop (add1 n) (add1 passed) discards)]
      [(discard) (loop (add1 n) passed (add1 discards))])))))
  \end{racketcodesmaller}
  \caption{Simple Generational Property Runner in Racket}
  \label{fig:simple-generational-testing-loop-racket}
\end{figure}

As users now have access to the building blocks for writing this loop, they are
empowered to change it! For instance, we have experimented with implementing a
precondition bypassing \roverlay{checker}{run} function that can be used to optimize
property execution for correct-by-construction generators by skipping the preconditions.
The users can change the shrinking algorithm, adjust sizes, move to structured printing,
or even write the results directly to a file. They can decide not to terminate the test
after a single counterexample, add specific budgets for number of discards or time bounds
if they decide to. All of these choices, and many more that we envision users will invent,
can be used to augment the effectiveness of testing.

\subsubsection{Mutation-Based Property Runners}

Property-Based Testing, for a long time, has relied solely on random
generation from scratch as opposed to mutating existing inputs as
fuzzing tools do. This design has been challenged in Targeted
PBT~\cite{TargetedPBT} with a mutational approach that optimizes
explicit user-provided feedback functions, and later by
FuzzChick~\cite{FuzzChick} and Zest~\cite{Zest}, which rely on branch
coverage information obtained via binary instrumentation, followed by
libraries such as HypoFuzz~\cite{HypoFuzz} and Bolero~\cite{bolero}.

\begin{figure}
  \noindent \begin{minipage}{0.62\textwidth}
    \begin{rocqcodesmaller}
      Definition fuzzLoop (fuel : nat) (cprop : Prop \emp) {Pool}
      $\roverlay{seedpool}{\tt  \{pool: SeedPool \}}$ (seeds : Pool) : G Result :=
      let fix fuzzLoop' (fuel passed discards: nat) seeds :=
      match fuel with
      | O => ret (mkResult discards false passed [])
      | S fuel' =>
      $\overlayfull{seedpool}{let directive := sample seeds in}$
      input <- match directive with
      $\overlayfull{generator}{| Generate => gen cprop (log2 (passed + discards))}$
      $\overlayfull{mutator}{| Mutate source => mutate cprop source}$
      end;;
      $\overlayfull{feedback}{ \overlayfull{checker}{res <- instrumentedRun cprop withInstrumentation;;} }$
      let '(res, feedback) := res in
      match res with
      | Normal seed false => (* Fails *)
      $\overlayfull{shrinker}{let shrunk := shrinkLoop 10 cprop seed in}$
      $\overlayfull{printer}{let printed := print cprop 0 shrunk in}$
      ret (mkResult discards true (passed + 1) printed)
      | Normal seed true =>  (* Passes *)
      $\overlayfull{seedpool}{match useful seeds feedback with}$
      | true =>
      $\overlayfull{seedpool}{let seeds' := invest (seed, feedback) seeds in}$
      fuzzLoop' fuel' (passed + 1) discards seeds'
      | false =>
      let seeds' := match directive with
      | Generate => seeds
      $\overlayfull{seedpool}{| Mutate \_ => revise seeds}$
      end in
      fuzzLoop' fuel' (passed + 1) discards seeds'
      end
      | Discard _ _ =>  (* Discard *)
      match directive with
      | Generate => fuzzLoop' fuel' passed (discards+1) seeds
      | Mutate source =>
      $\overlayfull{seedpool}{match useful seeds feedback with}$
      | true =>
      fuzzLoop' fuel' passed (discards+1) seeds
      | false =>
      fuzzLoop' fuel' passed (discards+1) $\overlayfull{seedpool}{(revise seeds)}$
      end
      end
      end
      end in
      fuzzLoop' fuel 0 0 seeds.
    \end{rocqcodesmaller}
    \captionsetup{justification=raggedleft}
    \captionof{figure}{Coverage-Guided Fuzzing Property Runner in Rocq}
    \label{fig:fuzzloop}
  \end{minipage}
  \begin{minipage}{0.37\textwidth}

    The coverage-guided fuzzing property runner introduces some complexity on top
    of the simple QuickCheck-style runner. This complexity is mainly related
    to the orchestration logic that manages the seed pool, which is parametric
    over the \textcolor{seedpool}{\tt SeedPool} interface. At each iteration of
    the loop, the seed pool produces a directive, either to generate an input
    from scratch, or mutate a previous input. The generated input is then passed
    into \roverlay{feedback}{\roverlay{checker}{\tt instrumentedRun}} function
    that is also parameterized over a custom instrumentation function.

    \vspace{4mm}

    In classic fuzz testing, this instrumentation function is branch or
    path coverage, yet our Rocq library can accommodate any function that
    observes information about the state of the executed program, as
    in \citet{FuzzFactory}.  This is reflected in the fuzzing loop in
    Fig. \ref{fig:fuzzloop} where feedback is received from the execution
    of the \roverlay{feedback}{\overlay{checker}{\tt instrumentedRun}}.
    Such feedback can range from traditional branch or path coverage (as
    in coverage-guided fuzzing) to timing or memory usage (as in
    performance fuzzing~\cite{PerfFuzz}).


    \vspace{4mm}

    This way, we view the fuzzing property runner presented in Fig.~\ref{fig:fuzzloop}
    as (1) a more generalized version of the classic coverage-guided fuzzing,
    and (2) a property-based testing version
    of FuzzFactory~\cite{FuzzFactory}, which allowed for arbitrary
    instrumentation functions to guide the search similar to the fuzzing
    property runner we present here.

  \end{minipage}
\end{figure}

An important design decision we must make when implementing a DBAS property encoding
in a statically typed language, such as our Rocq encoding, is which associated
functions are part of the property encoding. These recent advances in
mutational PBT have led us to make mutation a first class citizen of
our library, and we have implemented a generic seed pool interface
that mutation-based property runners can leverage. The seed pool interface
abstracts away search strategy (how to select which input to
mutate) and power schedule (how long to fuzz it for) concerns. We will
revisit this abstraction in just a few sections (Sec~\ref{sec:seedpool}).
We have implemented two property runners using these components:
a coverage-guided fuzzing property runner
as presented in Fig.~\ref{fig:fuzzing-property-runner}, and a custom feedback-guided
(targeted) property runner as presented in Fig.~\ref{fig:targeted-property-runner}.
We present the code of the fuzzing property runner in detail in Fig.~\ref{fig:fuzzloop},
and leave the presentation of the targeted property runner to the Appendix.






\subsection{Comparison of Deep and Shallow Embeddings}\label{sec:shallow_vs_deep}

In this section, we compare the performance of a DBAS-powered
implementation to its shallow counterpart. We turn to the
ETNA~\cite{ETNA} framework for evaluating property-based testing
performance, which comes with a series of Rocq workloads---programs
along with injected bugs---in the form of Binary Search Trees,
Red-Black Trees, and the Simply Typed Lambda Calculus. For the
purposes of this experiment we extended the ETNA tool to support
Racket and ported these workloads.

\begin{figure}[t]
  \includegraphics[scale=0.10]{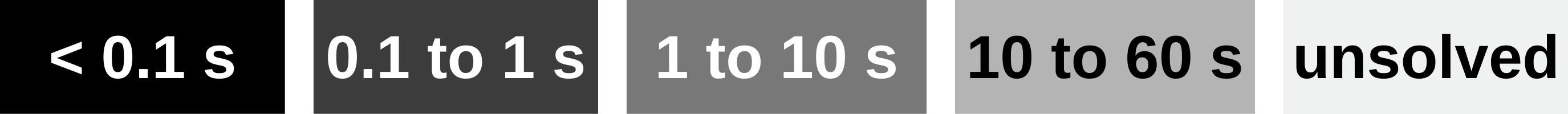}
  \caption{ETNA Style Bucket Chart Legend}\label{fig:etna-legend}
\end{figure}

Throughout the sequence of case studies that follow, our performance
results will use ETNA bucket charts: each bucket represents the tasks
(mutant-property pairs) solved within a certain time limit in the
average of 10 trial runs. The leftmost bucket with the darkest color
denotes the tasks solved within 0.1 seconds, where the remaining
buckets progressively denote the tasks solved within 1, 10, 60
seconds, and the last bucket denotes the tasks that were not solved
within 60 seconds for at least one of the 10 trial runs. The legend
for these charts is shown in Fig.~\ref{fig:etna-legend}.

\subsubsection{Comparison of Deep and Shallow Embeddings in Rocq}\label{sec:shallow_vs_deep-rocq}

Our first case study focuses on the performance implications of using
deferred binding abstract syntax instead of the standard
QuickCheck-style runner.

We benchmark our implementations of this loop for both Rocq and Racket
against the existing loops of the \qc{} and \rc{} libraries on 3 ETNA
workloads: binary search trees (BST), red-black trees (RBT), and the
simply-typed lambda calculus (STLC). Our results show that libraries
implemented via DBAS, using the runner we showed earlier in this
section, is on par with both \qc{} and \rc{}.

\begin{figure}[h!]
  \centering
  \begin{minipage}{.5\textwidth}
    \centering
    \includegraphics[width=.8\linewidth]{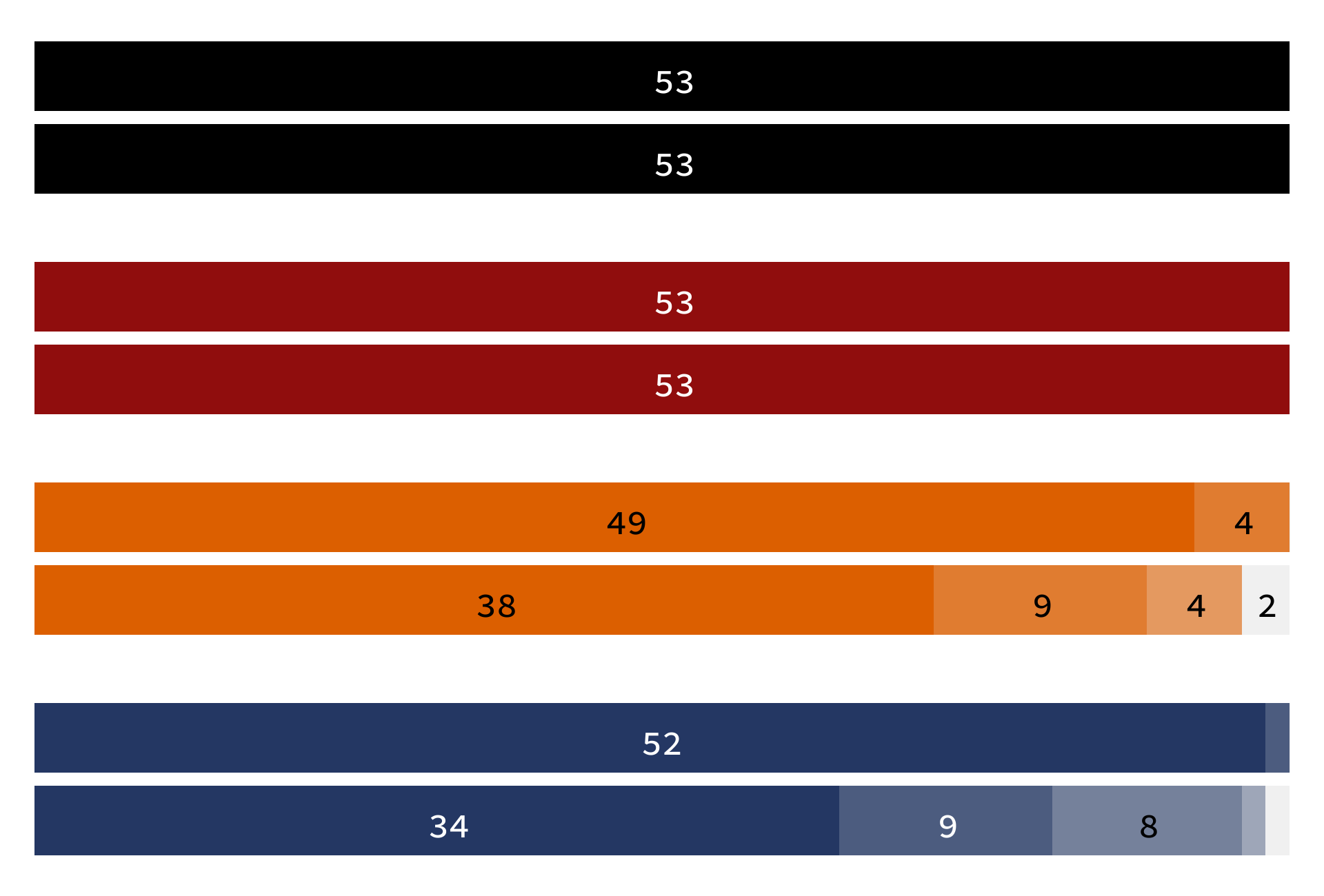}
    \caption{Binary Search Trees}
    \label{fig:bst_rocq}
  \end{minipage}%
  \begin{minipage}{.5\textwidth}
    \centering
    \includegraphics[width=.8\linewidth]{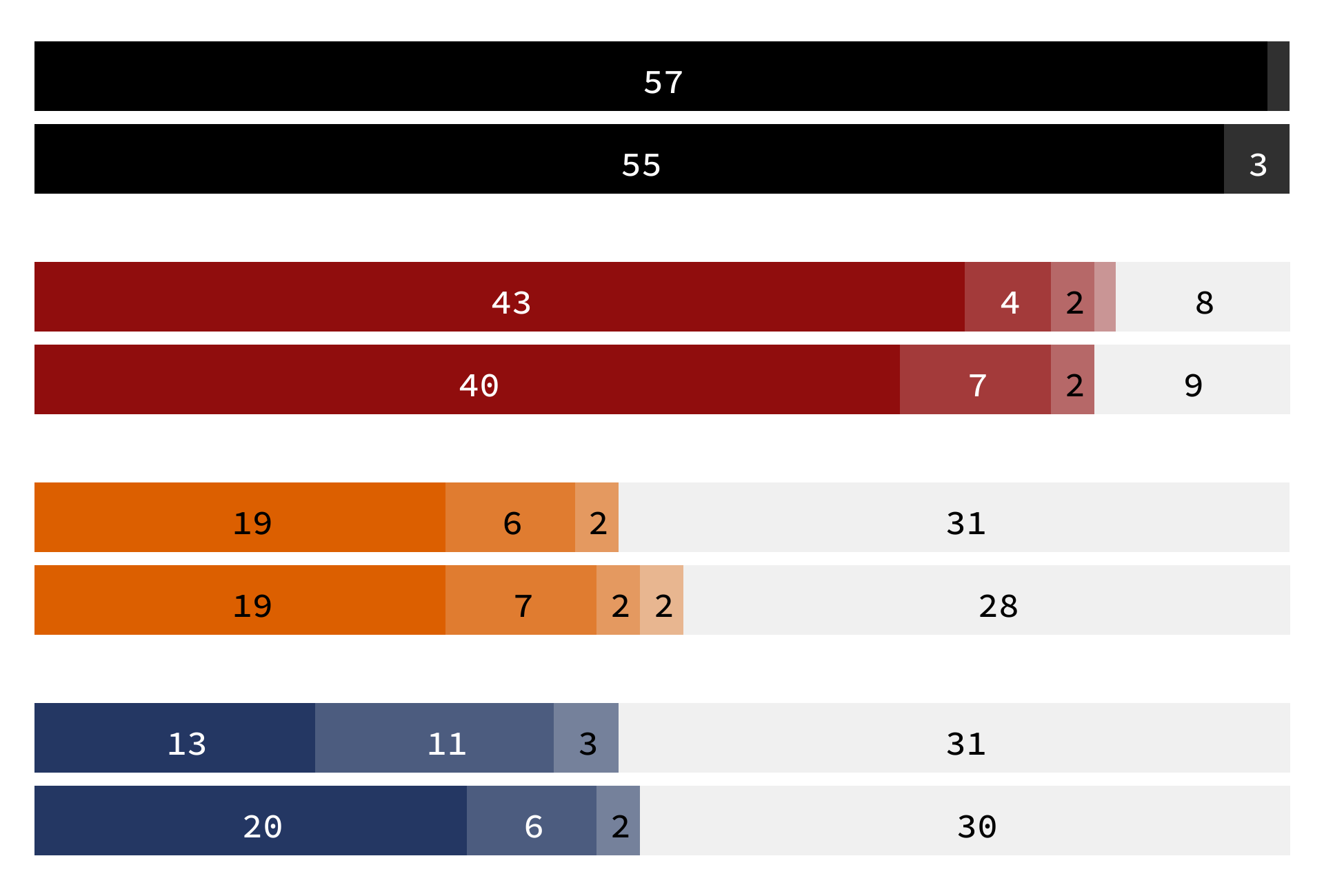}
    \caption{Red-Black Trees}
    \label{fig:rbt_rocq}
  \end{minipage}%

  \begin{minipage}{.6\textwidth}
    \centering
    \includegraphics[width=.8\linewidth]{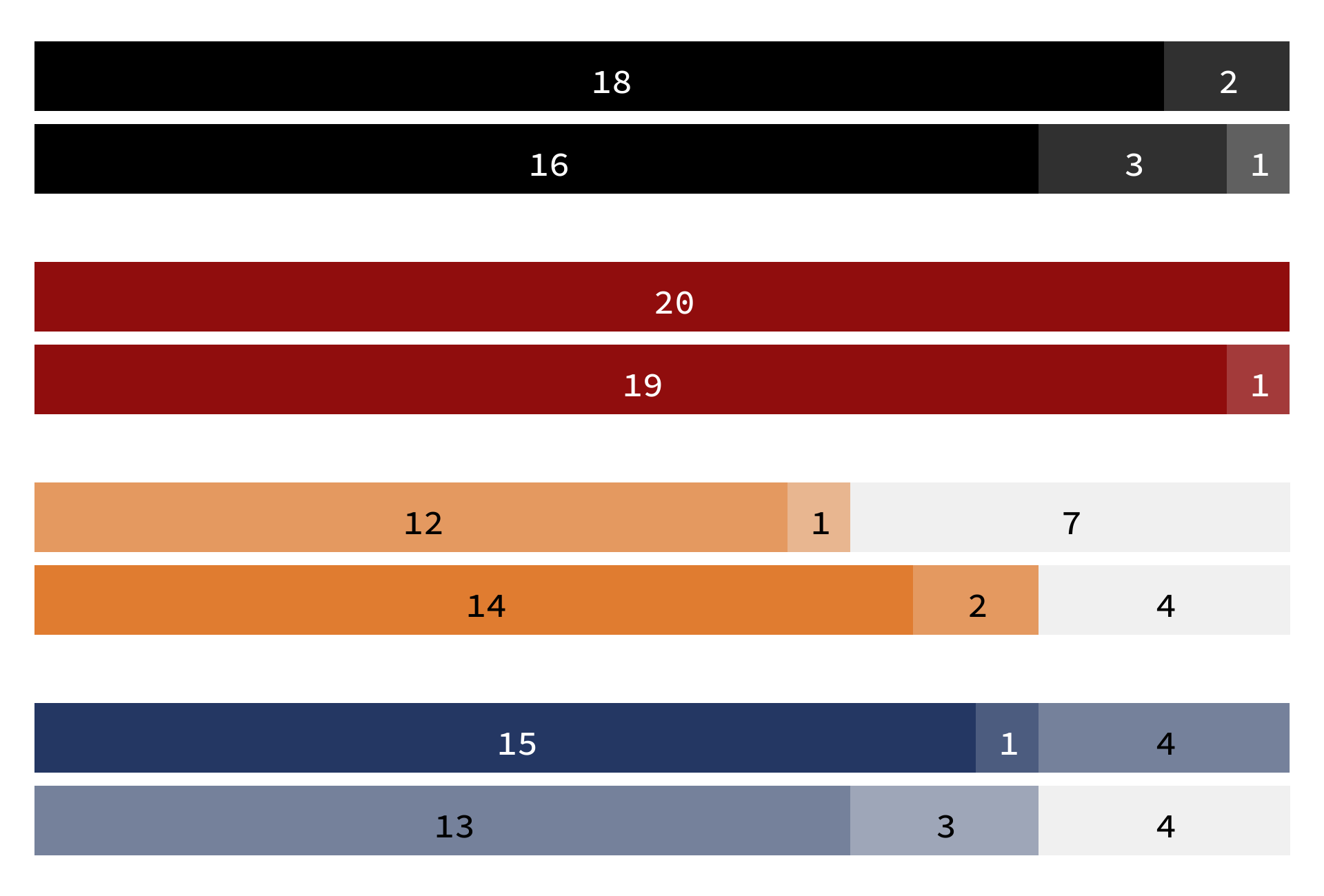}
    \caption{Simply-Typed Lambda Calculus}
    \label{fig:STLC_rocq}
  \end{minipage}%
  \caption{Comparison of Shallow and Deep Embeddings in Rocq.
    Buckets follow the legend in Fig.~\ref{fig:etna-legend}.
    Each color denotes a strategy, where the top bar is DBAS and the bottom bar is the shallow behavior.\\
    \explaincolor{chart_black}{Bespoke Generator},
    \explaincolor{chart_red}{Specification-Based Generator},\\
    \explaincolor{chart_orange}{Type-Based Fuzzer},
    \explaincolor{chart_blue}{Type-Based Generator}.
  }\label{fig:shallow_vs_deep_rocq}
\end{figure}

The bucket charts in Fig.~\ref{fig:shallow_vs_deep_rocq} show that
using DBAS does not incur a performance penalty compared to \qc{} in
the BST, RBT, and STLC workloads.  In a total of 12 strategy/workload
pairs, DBAS-style Rocq library outperforms \qc{} in 9 of them,
while \qc{} has a better performance in 3 of them. Yet, there are no
notable differences in the results of the two libraries in terms
of mean time to solve the tasks.

\subsubsection{Comparison of Deep and Shallow Embeddings in Racket}\label{sec:shallow_vs_deep-racket}

In our Racket experiments, we focused on comparing the DBAS-style
Racket library with the shallow embedding implemented
by \rc{}. We have conducted our experiments on the same BST, RBT, and
STLC workloads by porting the existing workloads in Haskell to
Racket. In the process of porting these workloads, we have also
discovered and reported a bug in the \rc{} core, which the authors
have fixed in the latest version of the library.
\footnote{The deanonymized version of our paper will have a citation to the bug report}

\begin{figure}[h!]
  \centering
  \begin{minipage}{.33\textwidth}
    \centering
    \includegraphics[width=.9\linewidth]{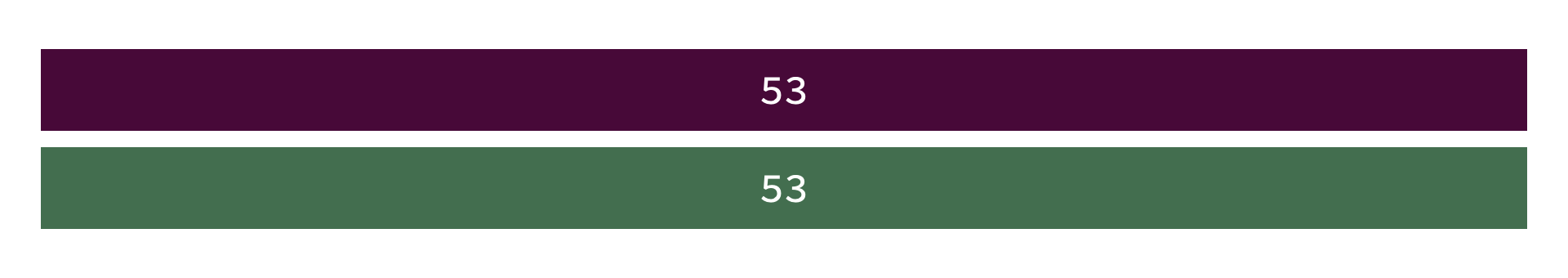}
    \caption{BST}
    \label{fig:bst_racket}
  \end{minipage}%
  \begin{minipage}{.33\textwidth}
    \centering
    \includegraphics[width=.9\linewidth]{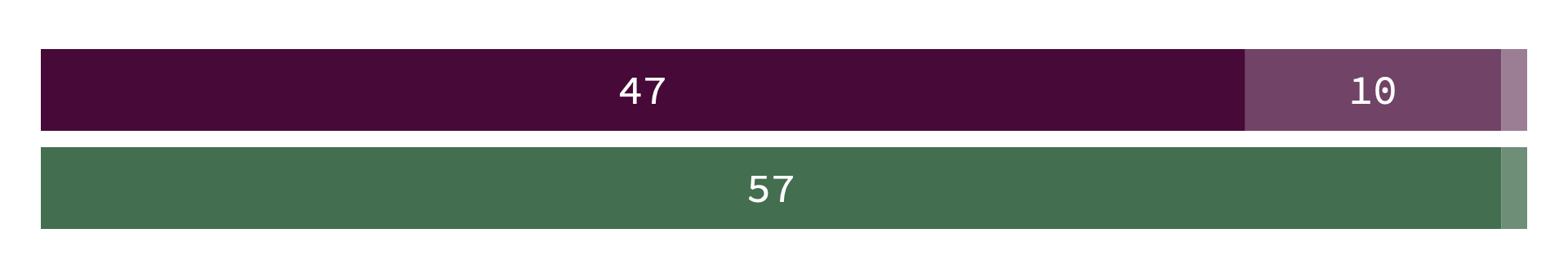}
    \caption{RBT}
    \label{fig:rbt_racket}
  \end{minipage}%
  \begin{minipage}{.33\textwidth}
    \centering
    \includegraphics[width=.9\linewidth]{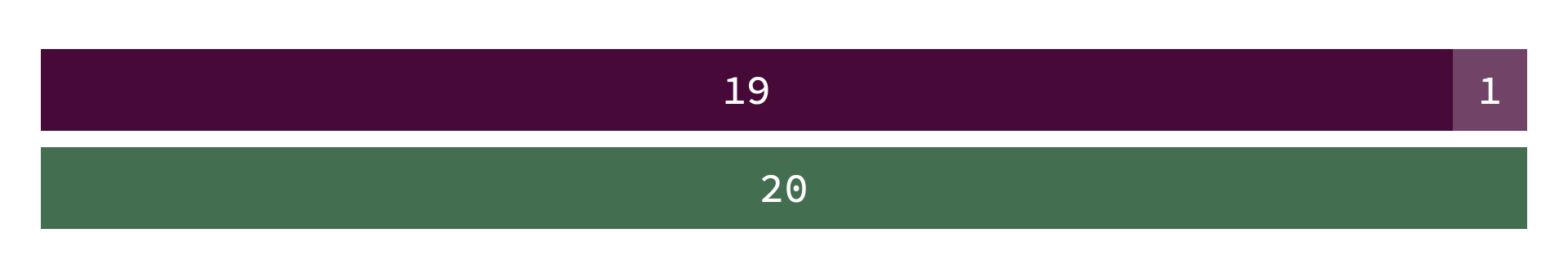}
    \caption{STLC}
    \label{fig:STLC_racket}
  \end{minipage}%
  \caption{Comparison of Shallow(\rc) and Deep(DBAS) Embeddings in Racket.
    Buckets follow the legend in Fig.~\ref{fig:etna-legend}.
    The purple bar on the top is DBAS with deep embedding, the green bar on the bottom is \rc{} with the shallow embedding.\\
    \explaincolor{chart_purple}{Bespoke Generator used with DBAS Library},
    \explaincolor{chart_green}{Bespoke Generator used with \rc{} Library}.\\
  }\label{fig:shallow_vs_deep_racket}
\end{figure}

Fig.~\ref{fig:shallow_vs_deep_racket} shows the results of the
comparison of the deep and shallow embeddings in Racket.  Once again,
we find no notable differences in performance between the two
versions. However, during the course of these experiments we found
that the default configuration of \rc{} in terms of size of generated
terms led to considerable performance degradation in the RBT case
study. As size is configurable in most PBT APIs, we have changed
the \rc{} size function to be logarithmic with respect to number of tests,
as our property-runner does. Still, this
further reinforces our point on programmability: if sizes and similar
aspects of generation are configurable (and severely impact testing
performance), why shouldn’t the runners themselves be configurable?


\subsection{DBAS-powered Experiments}

In this section, we carry out three case studies that the expressiveness
of DBAS enabled us to.

\subsubsection{An Exploration of Seed Pool Design Choices}\label{sec:seedpool}


First, we turn to feedback-guided property runners and their
programmability. Fuzzing research often explores different
strategies and power schedules~\cite{aflplusplus,
  10.1145/3551349.3559550,DBLP:conf/sigsoft/BohmeMC20,
  DBLP:journals/tse/BohmePR19}, with researchers coming up with better
and better designs. On the other hand, property-based testing
frameworks that support feedback-guided generation of inputs such as
FuzzChick or Hypothesis, pride themselves in the power of testing
arbitrary user-defined specifications, but do not provide an option to
configure such crucial parameters of their feedback-guided property
runners. We illustrated such a feedback-guided runner earlier in
this section, that abstracts away such concerns into a small API, which
we implement in Rocq using typeclasses
(Fig.~\ref{fig:seedpool_rocq_impl}).
\begin{figure}[h!]
  \begin{minipage}{.50\textwidth}
    \begin{rocqcodesmaller}
      Class SeedPool {A F Pool: Type} := {
          (* Creates an empty pool. *)
          mkPool  : unit -> Pool;
          (* Adds a useful seed into the pool. *)
          invest  : (A * F) -> Pool -> Pool;
          (* Decreases the energy of a seed after
          a useless trial. *)
          revise  : Pool -> Pool;
          (* Samples the pool for an input. *)
          sample  : Pool -> @Directive A F;
          (* Returns the best seed in the pool. *)
          best    : Pool -> option (@Seed A F);
        }.
    \end{rocqcodesmaller}
  \end{minipage}%
  \begin{minipage}{.50\textwidth}

    \begin{rocqcodesmaller}
      Class Utility {A F Pool: Type}
      `{SeedPool A F Pool} := {
          (* Returns true if the feedback
          is interesting. *)
          useful  : Pool -> F -> bool;
          (* Returns a metric of how interesting
          the feedback is. *)
          utility : Pool -> F -> Z;
        }.
    \end{rocqcodesmaller}
  \end{minipage}%
  \caption{SeedPool and Utility typeclasses used in Feedback-Guided Property Runners}
  \label{fig:seedpool_rocq_impl}
\end{figure}

The configurability enabled by DBAS allows the users to rely on
a set of community-accepted defaults chosen by framework developers, but also to explore
if different choices from the literature or novel ones they devised
fit their testing needs better. In this case study, we explore six
different data structure representations to hold interesting seeds, as
well as four different power schedules, leading to a total of 21
different configurations. We picked these configurations to explore
different parts of the design space, such as the queuing strategy, the
size/cardinality of the pool, whether the pool is monotonic, and how
many times a given seed is reused. We have conducted our experiments
on the IFC workload in ETNA, and we have used a type-based generator
alongside a type-based mutator to conduct our experiments. While our
experiment reveals a clear winner for this case study, we reiterate
that our primary goal is not the exploration itself, but rather to
demonstrate that such an exploration can be expressed by varying reusable
runner components without modifying the property language or framework internals.

More concretely, we explore the following data structure
representations, three that hold collections of seeds (as in
FuzzChick), and three that only hold a single seed (as in Targeted
PBT):
\begin{enumerate}
  \item {\em FIFO Queue Seed Pool}: A pool that holds a
        queue of seeds, and reduces the energy of the current seed after
        usage.  The pool only generates a new seed from scratch when the
        queue is empty, and mutates the seed otherwise. When the energy of
        the current seed drops to 0, it is removed from the queue. The next
        seed is chosen from the front of the queue. This was the default
        behavior of FuzzChick.

  \item {\em FILO Queue Seed Pool}: The same as the {\em FIFO Queue
            Seed Pool}, but the next seed is chosen from the back of the queue.

  \item {\em Heap Seed Pool}: Similar to the FIFO and FILO Queues, but
        the seeds are stored in a heap, creating a priority queue.

  \item {\em Static Singleton Pool}: A pool that holds a single seed,
        and does not reduce its energy after usage.  The pool generates a
        new seed from scratch at the first iteration, and mutates its seed
        for the subsequent iterations.  The seed is only updated when a new
        seed with a better feedback is generated via mutation. This
        essentially devolves the search to hill climbing, as in the original
        Targeted PBT work~\cite{TargetedPBT}.

  \item {\em Dynamic Monotonic Singleton Pool}: A pool that holds a single
        seed, and reduces its energy after usage.  The pool generates a new
        seed from scratch at the first iteration and when the energy of the
        current seed is 0, mutates the seed otherwise. The seed is only
        updated when a new seed with a better feedback is generated.

  \item {\em Dynamic Resetting Singleton Pool}: A pool that holds a single
        seed, and reduces its energy after usage.  As opposed to the {\em
            Dynamic Monotonic Singleton Pool}, once the current seed's energy is
        0, the seed is effectively discarded and a new seed is generated
        from scratch.

\end{enumerate}

All of the queues except the {\em Static Singleton Pool} have been
tested with 4 different energy schedules, where the energy of a new seed
was respectively up to 1, 10, 100, 1000, depending on its
interestingness. We report the experiments over 5 trials for each
configuration within 65 tasks in the IFC workload in ETNA. For brevity,
the graphs omit 34 tasks none of the configurations have solved, and
only show the remaining 31 tasks. Each bucket represents the tasks
solved within a certain time limit for at least one of the 5 trials,
where the leftmost bucket with the darkest color denotes the tasks
solved within 0.1 seconds, where the remaining buckets progressively
denote the tasks solved within 1, 10, 60 seconds, and the last bucket
denotes the tasks that were not solved within 60 seconds.

\begin{figure}[h!]
  \centering
  \begin{minipage}{.33\textwidth}
    \centering
    \includegraphics[width=.95\linewidth]{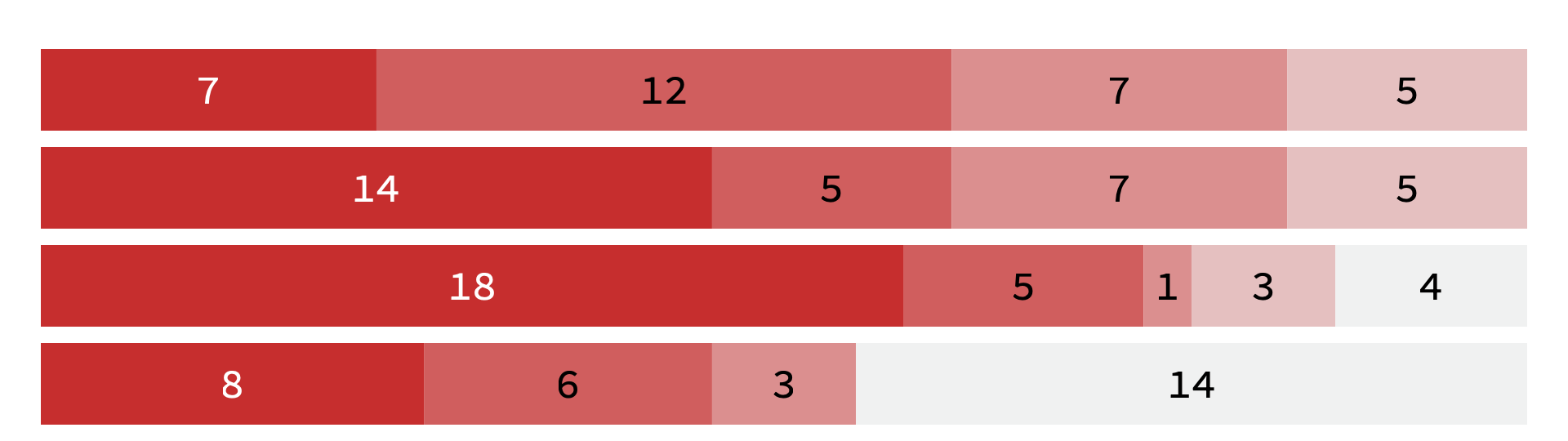}
    \caption{Heap Seed Pool}
    \label{fig:heapseedpool}
  \end{minipage}%
  \begin{minipage}{.33\textwidth}
    \centering
    \includegraphics[width=.95\linewidth]{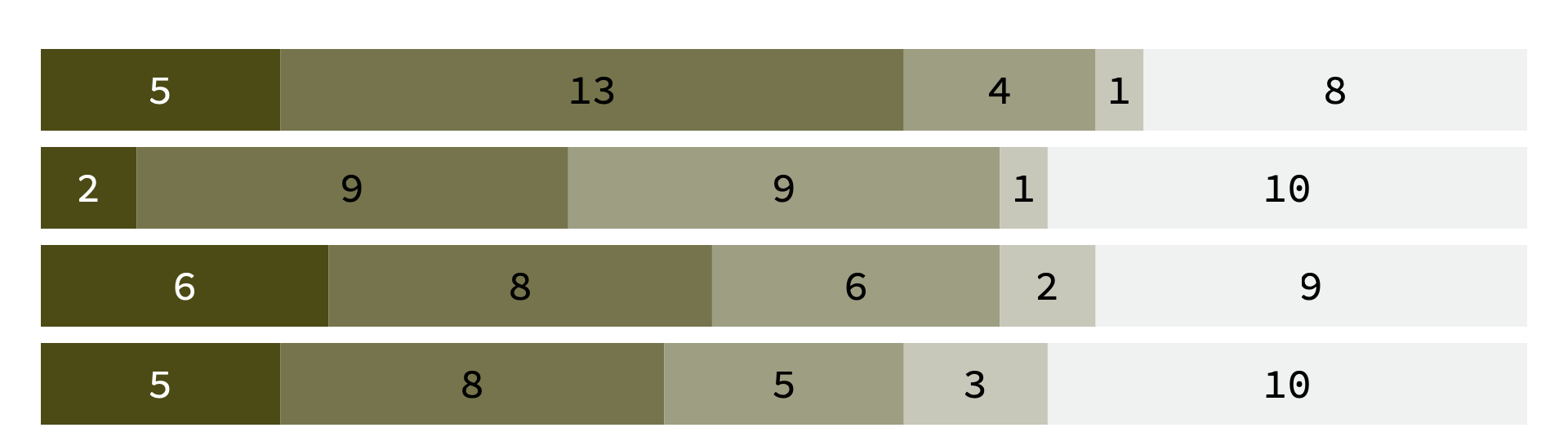}
    \caption{FILO Queue Seed Pool}
    \label{fig:filoseedpool}
  \end{minipage}%
  \begin{minipage}{.33\textwidth}
    \centering
    \includegraphics[width=.95\linewidth]{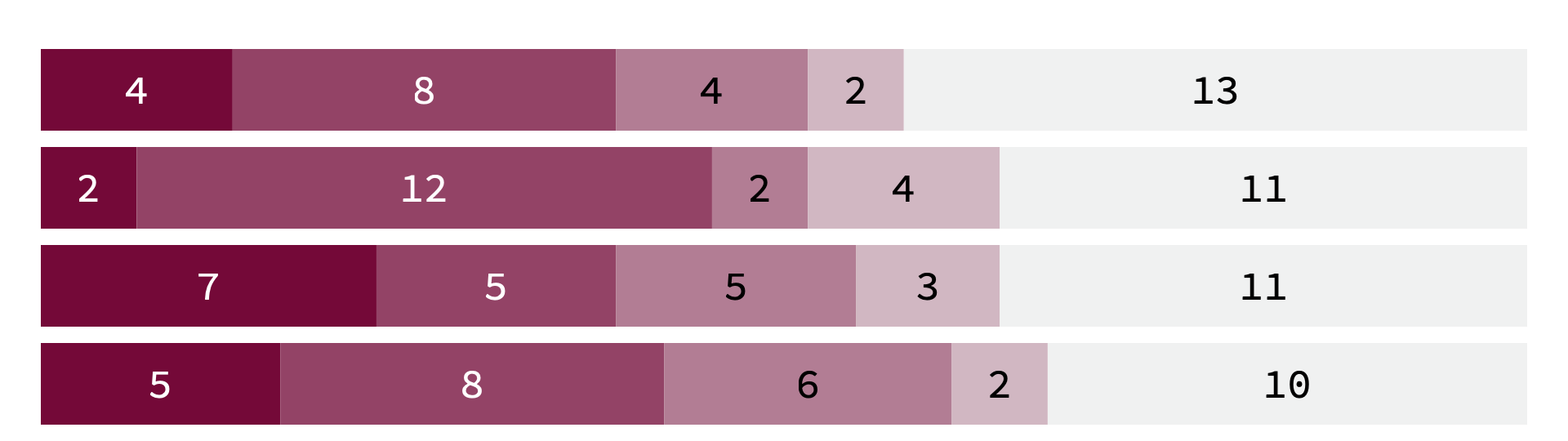}
    \caption{FIFO Queue Seed Pool}
    \label{fig:fifo_seedpool}
  \end{minipage}%

  \begin{minipage}{.33\textwidth}
    \centering
    \includegraphics[width=.95\linewidth]{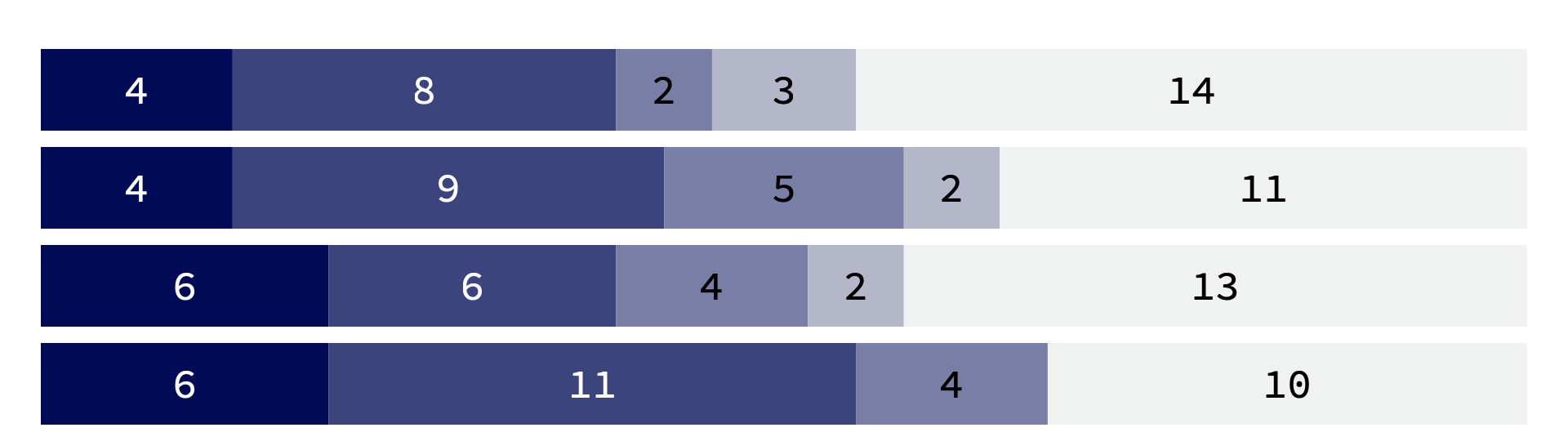}
    \caption{Dynamic Monotonic  \\Singleton Seed Pool}
    \label{fig:dynamicmonotonicseedpool}
  \end{minipage}%
  \begin{minipage}{.33\textwidth}
    \centering
    \includegraphics[width=.95\linewidth]{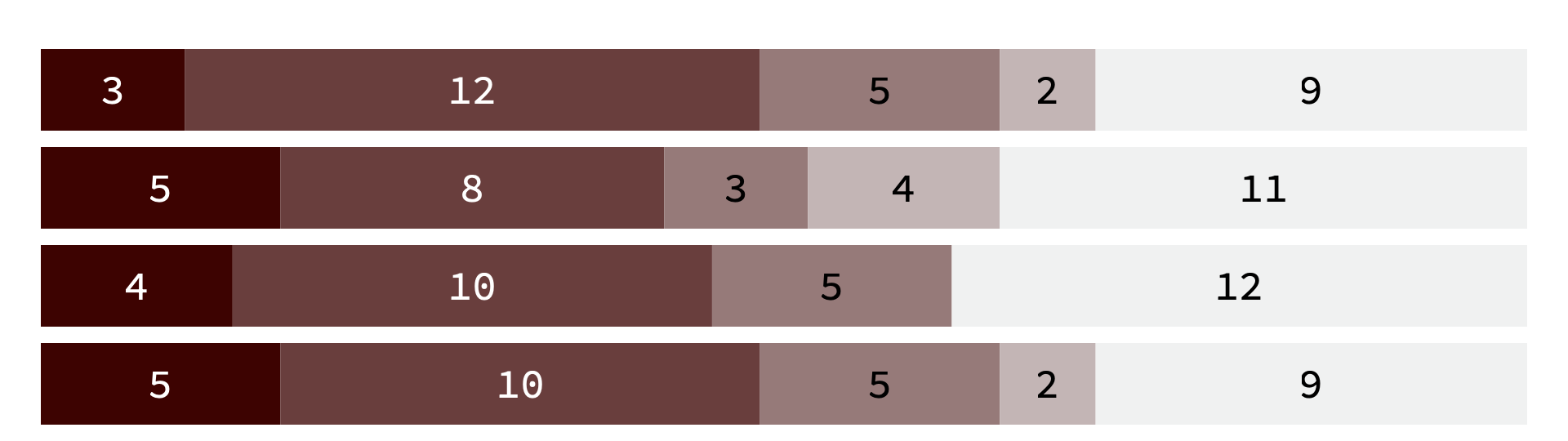}
    \caption{Dynamic Resetting \\ Singleton Seed Pool}
    \label{fig:dynamicresettingseedpool}
  \end{minipage}%
  \begin{minipage}{.33\textwidth}
    \centering
    \includegraphics[width=.95\linewidth]{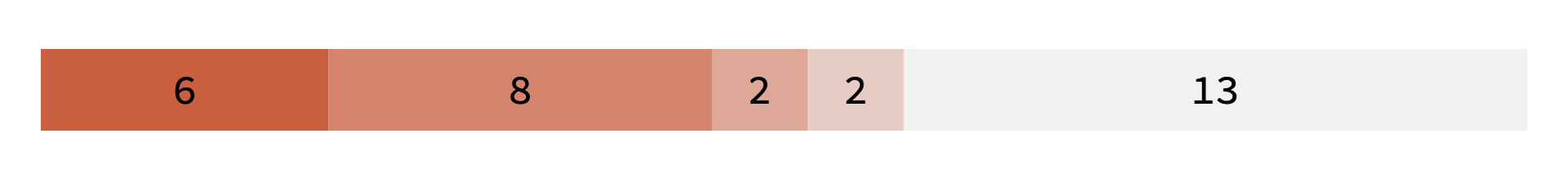}
    \caption{Static Singleton \\ Seed Pool}
    \label{fig:staticsingletonpool}
  \end{minipage}%
  \caption{Comparison of Seedpool Strategies in Rocq.
    Buckets follow the legend in Fig.~\ref{fig:etna-legend}.
    Each color denotes a strategy, where the bars are ordered from energy levels 1, 10, 100, and 1000;
    except for the static singleton pool that ignores energy.\\
  }\label{fig:seedpool_rocq}
\end{figure}

These results mark a task as ``solved'' for the purposes of a bucket
chart if at least 1 out of 5 fuzzing campaigns finds a bug. If we switched
to requiring {\em all} fuzzing campaigns to find the bug, the results paint
an even more compelling argument: {\em only} Heap-based pools consistently
find counterexamples---we omit the graphs because all other options fail
to consistently solve a task across runs, indicating a very high variance
that heap seed pool does not exhibit. As a result, we plan to submit a PR to QuickChick
to use the effective configuration as a default going forward.

\subsection{Comparison of Integrated Shrinking with External Shrinking}\label{sec:shrink-comp}

Test case reduction, or counterexample minimization, or shrinking, is a fundamental
aspect of random testing~\cite{zeller2002simplifying}, however it garners much
less attention than its random generation counterpart. A major splitting
point in PBT frameworks has been their approach to shrinking, with the
original QuickCheck using delta-debugging style structural shrinking where
the generated counterexample structure is shrunk by removing
its substructures to produce smaller versions with an accompanying
reproducer that tests if the smaller versions keep triggering
the bug; and Hypothesis~\cite{HypothesisShrinking} in Python, falsify~\cite{falsifyLeo2023}
in Haskell, and \rc~\cite{Rackcheck} in Racket use an alternative route,
instead of shrinking
the generated structure, they shrink the source of randomness that led to the
generation. This method, called integrated or internal shrinking, removes
the requirement of a separate shrinking function and preserves the generation
invariants in the generator while making no guarantees on the similarity
of the shapes obtained in the shrinking phase.

We argue, as we have done many times in the paper, that such splitting
points should not be decisions at the level of the library, but rather
decisions to be made in relevant contexts by the users of the library.
In an ideal scenario, a user should be able to switch between using an
integrated shrinker versus an external one, and vice versa; and perhaps
even use or build a new approach that might not be available when the
library was written in the first place. Instead of forcing the users
to choose or change libraries based on their context and approach to shrinking,
DBAS allows for building your own property runner with its custom approach
to shrinking, and that is precisely what we have done in this experiment.

We have compared the effectiveness of the integrated shrinker
of \rc{} against a simple external shrinker we have implemented
in our DBAS PBT library in Racket. We have conducted our experiment
on the {\tt System F} workload in ETNA, and we used the same generator in both
\rc{} and DBAS-style Racket library, equipping our library with a simple type-based external shrinker we
implemented in place of the integrated shrinker.

\begin{figure}[h!]
  \centering
  \begin{minipage}{.5\textwidth}
    \centering
    \includegraphics[width=0.9\linewidth]{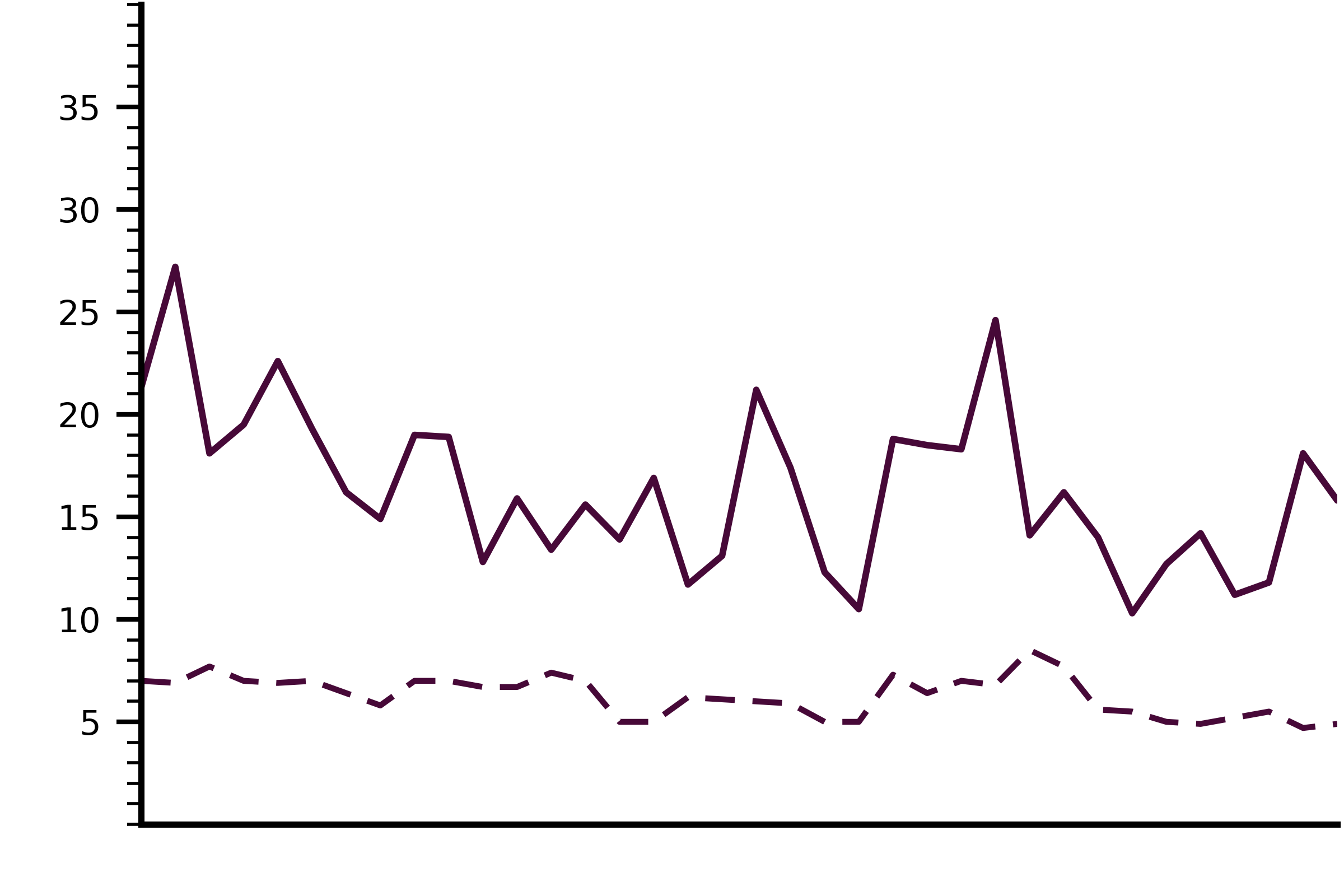}
  \end{minipage}%
  \begin{minipage}{.5\textwidth}
    \centering
    \includegraphics[width=0.9\linewidth]{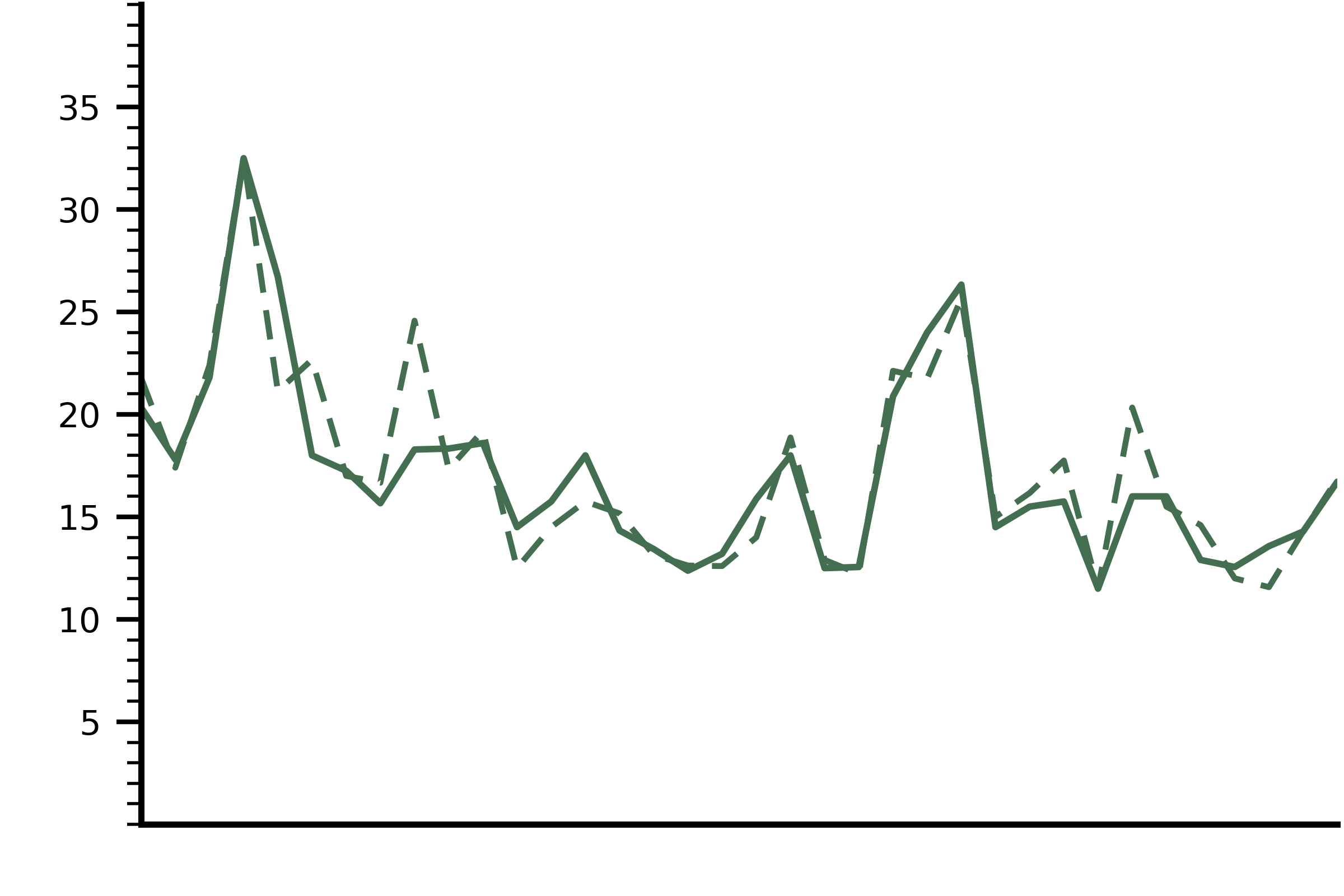}
  \end{minipage}%
  \caption{Original sizes (continuous) and shrunk sizes (dashed) across tasks (x axis) for
    the external shrinker (left) and internal shrinker (right).
  }\label{fig:shrinking_line}
\end{figure}

Our results show that the external shrinker
successfully shrinks System F terms to a minimal counterexample that is an
average of 2.66 times smaller than the original input with a standard deviation
of 1.23, while the integrated shrinker
of \rc{} only shrinks the inputs to an average rate of 1.04 times smaller than the
original input with a standard deviation of 0.37.
\rc{} only shrunk the inputs
to smaller inputs in 66 out of the 360 trials, kept the size the same in 77, grew
the inputs in 68, and failed to shrink in 149 trials.
Fig.~\ref{fig:shrinking_line} depicts these results, which might be
surprising at first glance: does the internal shrinker really only
shrink the inputs in 20\% of the trials? It turns out that in this
case, it does. Its notion of size is based on
the structure of the generator rather than on the input itself, so
"smaller randomness" may not actually lead to smaller input
values.

The point of this experiment is not to demonize internal shrinking---for many
testing situations, it is perfectly sufficient and much more user-friendly than
a bespoke shrinker. However, in pathological cases, programmers need an escape hatch,
and DBAS provides the necessary configurability.

%

\subsection{Parallelizing Property-Based Testing} \label{sec:parallel-experiment}

An important recent work on novel property runners is QuickerCheck~\cite{ParallelTesting},
where authors implement and evaluate a parallel run-time for QuickCheck~\cite{ClaessenH00},
achieving massive performance gains on a variety of workloads. The underlying idea the authors
propose is rather simple, they provide an alternative, parallel property runner {\tt quickCheckPar},
where multiple worker threads share a common variable that controls the size of the generated inputs.
Using this strategy, the authors achieve an almost linear speed-up with respect to the number of physical
cores used in testing. Unfortunately, due to the rigid structure of shallow embedding based PBT libraries,
creating such a parallel runner requires intensive engineering effort. The commit history of the
project shows that the implementation of this new runner is an effort spanning 2 years and more than 50
commits.

In contrast, the DBAS-style Racket library let us implement a naive version of
the QuickerCheck algorithm as a user-space runner, without changing the property
representation or the core framework.
Our implementation uses worker threads
with 2 shared atomic variables, one keeping the current number of tests across threads, and one
indicating if the process of testing is finished or not.

We have compared this parallel runner with 4 threads
against the single threaded runner across 131 tasks (53 BST, 58 RBT, 20 STLC).
As the results of the simple generational property runner evaluation in Racket (\S~\ref{sec:shallow_vs_deep-racket})
shows, the majority of the tasks in ETNA~\cite{ETNA} are trivial for bespoke generators, they can be
solved within 0.1 seconds. In order to measure the impact of the parallelization in the presence
of such variation across tasks, we apply a set of cutoffs of time differences between task solving performance,
and report the average ratio of time to solve for each cutoff.

\begin{itemize}
  \item For all tasks where the difference between time-to-failure for single and
        parallel runners is greater than a millisecond, the average of their ratios is 1.2,
        where the parallel runner takes 20\% more time {\it on average} than the single threaded
        runner across 49 of the 131 tasks.
  \item For all tasks where the difference between time-to-failure for single and
        parallel runners is greater than 0.1 seconds, the average of their ratios is 0.33,
        where the parallel runner is {\it on average} 3 times faster than the single threaded
        runner across 10 of the 131 tasks.
  \item Only 1 task has a difference of more than 1 second, which the average time-to-failure
        for single threaded runner is 3.56 seconds, and the average time-to-failure for the
        parallel runner is 1.16 seconds, resulting in again a 3 times speed-up due to parallelization.
\end{itemize}

\section{Related Work}\label{sec:related}

Throughout this paper, we have thoroughly discussed various
property-based testing frameworks, their property languages, and the
property runners that they come bundled with. Here, we briefly summarize
related work in PBT and we also discuss related work in deeply
embedded domain-specific languages.

\paragraph{Property-Based Testing Frameworks}

To our knowledge, no widely-used frameworks expose a reified property representation
that supports user-defined runners without modifying library internals. However, there are
multiple frameworks that make distinctly different choices in what
capabilities they provide to users. Table~\ref{tab:frameworks}
summarizes the status quo in popular tools.

\begin{table}[h]
  \begin{center}
    \begin{tabular}{ c|c|c|c}
      \hline
      {\bf     Framework}
                                     & \raggedright {\bf Language  }
                                     & \raggedright {\bf Shrinking }
                                     & {\bf Feedback  }                                                                                      \\
      \hline
      \hline
      QuickCheck~\cite{ClaessenH00}  & Haskell                       & External                            &                                 \\
      \hline
      Hedgehog~\cite{Hedgehog}       & Haskell                       & Internal                            &                                 \\
      \hline
      QuickChick~\cite{Foundational} & Rocq                          & External                            & Coverage~\cite{FuzzChick}       \\
      \hline
      Hypothesis~\cite{Hypothesis}   & Python                        & Internal~\cite{HypothesisShrinking} & Coverage$^*$                    \\
      \hline
      Zest~\cite{Zest}               & Java                          & AFL Trimming                        & Customizable~\cite{FuzzFactory} \\
      \hline
      QCheck                         & OCaml                         & Both$^*$                            &                                 \\
      \hline
      Crowbar                        & OCaml                         & AFL Trimming                        & Coverage                        \\
      \hline
      RackCheck                      & Racket                        & Internal                            &                                 \\
      \hline
      QuviQ QuickCheck               & Erlang                        & Both                                &                                 \\
      \hline
      PropEr                         & Erlang                        & Both                                & Targeting~\cite{TargetedPBT}    \\
      \hline
    \end{tabular}
  \end{center}
  \caption{Shrinking and feedback options in selected set of widely used frameworks; $^*$ denotes
    experimental or partial support.}
  \label{tab:frameworks}
\end{table}

\paragraph{Free Generators}

The design of our deeply embedded property language builds on a rich
literature of embedded
DSLs~\cite{hudakBuildingDomainspecificEmbedded1996}. In particular,
our approach parallels work on {\em free
    generators}~\cite{goldsteinParsingRandomness2022}, which present a
deeply embedded DSL for writing random generators. Deeply embedded
properties and generators are largely orthogonal---free generators may
be able to simplify the implementation of some of the different
runners described in \S~\ref{sec:eval}, but they do not allow the
developer to re-program the loop itself.
This follows a more general trend of using free structures to increase
expressivity or usability in the programming languages community. For
example, itrees~\cite{DBLP:journals/pacmpl/XiaZHHMPZ20} introduced a general-purpose Rocq structure
which is essentially a coinductive variant of free monads, which
allows them to represent and reason about interactive recursive
programs. In follow up work, \citet{DBLP:journals/pacmpl/LiW22}
explored how other free structures (such as applicative functors)
can be used as an alternative to free-monad-based embeddings.

\paragraph{Mixed Embeddings}
There is also a long line of related work in attempting to bridge the
benefits of shallow embeddings (ease of use, as in the current
property language) with those of deep ones (extensibility). For
example, \citet{CARETTE_KISELYOV_SHAN_2009} showed how to use
typeclasses in Haskell to allow for shallow embeddings that can be
interpreted in different ways, hinting at a way of incorporating a
deeper property language in a type system such as Haskell's.  More
recently, \citet{Unembedding} showed how to convert between embedding
representations by {\em unembedding} alleviating some of the problems
with Higher-Order Abstract Syntax representations.
Finally, \citet{DeeperShallowEmbeddings} introduced a hybrid embedding
where typing derivations are represented as a deep embedding indexed
by shallow terms in the host language, offering pattern matching
capabilities.


\section{Conclusion and Future Work}\label{sec:future}

We have presented Programmable Property-Based Testing, a novel PBT paradigm
that changes the underlying representation of the properties using a mixed
embedding, deferred binding abstract syntax (DBAS), we introduce. Programmable PBT
is a new approach to encoding properties for PBT that
enables more flexible and programmable testing. The key advance made by DBAS
is to reify properties as a free data structure; allowing them to be written
in a clear and readable way, separate from the property runner that tests them.
These more deeply embedded properties can then be inspected and interpreted by
user-defined property runners. With the help of DBAS, developers in Rocq, Racket,
and hopefully soon other programming languages, can tailor, experiment, and
adapt their testing setup to their domain.

In the future, we intend to make DBAS convenient to program in more
mainstream languages than Rocq and Racket.
On the static typing side, we chose to implement our approach in Rocq
as its dependent type system and powerful notation mechanism allowed
us to conveniently express the heterogeneous contexts involved. The
same approach should also be implementable in Haskell using standard
GADT-based heterogeneous lists~\cite{HaskellPromotion}; however, without access to notational
conveniences as in Rocq, it remains future work to see how to regain a
similar degree of ergonomics in that setting.
On the dynamic typing side, the techniques described should be readily
transferrable to other languages. For example, given Python's powerful
reflection capabilities and its convenient decorator features, a
deferred-binding style property language should be implementable on top
of, e.g., Hypothesis.
%
%

\section{Data Availability Statement}
All our work will be made publicly available. We have submitted
an artifact for artifact evaluation that includes the implementations of
the property language in Rocq and Racket, as well as scripts to
re-execute the experiments carried out. Once deanonymized, we
will add the artifact URL to the camera-ready version of this paper.

\bibliographystyle{ACM-Reference-Format}
\bibliography{quick-chick,leo,harry}

\newpage

\appendix

\section{Appendix} \label{sec:appendix}

\subsection{Custom-Feedback Guided (Targeted) Property Runner} \label{sec:targeted-testing-loop}

In recent years, PBT tools and mutation based fuzzers have begun to find common ground.
On one hand,
fuzzing tools have been trying to move towards more and more
structured generation of inputs as well as incorporate and encode more
complex properties than simply ``the program doesn't
crash''~\cite{DBLP:journals/pacmpl/ParkWZS21,DBLP:conf/sigsoft/Sun0JWP023,DBLP:conf/icse/PadhyeLSPT19}.
On the other hand, we have seen a rise in the popularity of
property-based testing tools that are able to guide the generation of
inputs using feedback~\cite{TargetedPBT, FuzzChick, Zest}.
Following this trend, we have used DBAS to implement mutation-based generation,
developing two property runners leveraging mutation and feedback.

A mutation-based targeted property runner has two important differences
from the simple generational property runner described in
Fig.~\ref{fig:simple-generational-testing-loop-rocq}: the feedback and the targeting.
It uses a genetic algorithm based on a user provided custom feedback function for
guiding the search towards interesting inputs, and it employs
a user-provided or type-derived mutator functions to mutate the input
it currently focuses on.  The main benefit of such a runner is that it means
developers can potentially avoid writing
complex generators for complex types and preconditions.

The runner is further parameterized by a seed pool (to keep track of
interesting inputs) and a utility function (to accommodate different
types of feedback). As shown in the pioneering work
of \citet{FuzzFactory}, customizable
feedback under user control can lead to very effective testing,
and DBAS allows even more relevant choices to be made without
needing to modify the internals of a framework. A pictorial depiction
of the targeted property runner
is illustrated in Fig.~\ref{fig:targeted-property-runner}.

In Fig.~\ref{fig:targetloop}, we provide an annotated Rocq implementation of the
custom-feedback guided (targeted) property runner in Rocq. Similar to the
simple generational runner or
coverage-guided (fuzzing) runner in \S~\ref{sec:eval},
this runner reuses the building
blocks we have provided for writing novel property runners.

It is important to note that this specific runner is not a fixed part of our library, but merely a default
behavior that is convenient and can be readily used. Alternative implementations may change the feedback behavior
to accommodate feedback in the discard cases, or change the printing or shrinking behaviors. Even the seed
pool and utility typeclasses are provided as sensible defaults and building blocks, rather than static
design choices as is the case in shallow embedding based PBT frameworks.

\subsection{Parallel Property Runner}\label{sec:parallel-loop-code}

In Fig.~\ref{fig:par-loop-appendix} is a slightly abridged version of our Racket parallel testing runner.
Parallelism is done through Racket futures. We use a lock-free shared counter
and a flag that is set if a counterexample is found, which is necessary because
Racket futures are not able to be halted arbitrarily. Each worker thread grabs
a test number from the counter, and loops until it either finds a
counterexample or the counter exceeds the test number, with the main thread
waiting for results from the workers and combining them when they finish.

Like all the other loops we introduced, this loop is not a fixed part of the library, and more efficient or
sophisticated parallel runners can be implemented without changing the
underlying property representation.

\begin{figure}

  \noindent \begin{minipage}{0.60\textwidth}
    \begin{rocqcodesmaller}
      Definition targetLoop (fuel : nat) (cprop : Prop \emp)
      $\textcolor{feedback}{\tt (feedback\_function:\llbracket \{ cprop \} \rrbracket -> Z)}$ {Pool : Type}
      $\textcolor{seedpool}{\tt  \{pool: SeedPool \}}$ (seeds : Pool)
      $\textcolor{seedpool}{\tt (utility: Utility)}$ : G Result :=
      let fix targetLoop' (fuel : nat)
      (passed : nat) (discards: nat)
      (seeds : Pool) : G Result :=
      match fuel with
      | O => ret (mkResult discards false passed [])
      | S fuel' =>
      $\overlayfull{seedpool}{let directive := sample seeds in}$
      input <- match directive with
      $\overlayfull{generator}{| Generate => gen cprop (log2 (passed + discards))}$
      $\overlayfull{mutator}{| Mutate source => mutate cprop source}$
      end;;
      $\overlayfull{checker}{res <- run cprop input;;}$
      match res with
      | Normal seed false => (* Fails *)
      $\overlayfull{shrinker}{let shrunk := shrinkLoop 10 cprop seed in}$
      $\overlayfull{printer}{let printed := print cprop 0 shrunk in}$
      ret (mkResult discards true (passed + 1) printed)
      | Normal seed true =>  (* Passes *)
      $\overlayfull{feedback}{let feedback := feedback\_function seed in}$
      $\overlayfull{seedpool}{match useful seeds feedback with}$
      | true =>
      $\overlayfull{seedpool}{let seeds' := invest (seed, feedback) seeds in}$
      targetLoop' fuel' (passed + 1) discards seeds'
      | false =>
      let seeds' :=
      match directive with
      | Generate => seeds
      $\overlayfull{seedpool}{| Mutate source => revise seeds}$
      end in
      targetLoop' fuel' (passed + 1) discards seeds'
      end
      | Discard _ _ => (* Discard *)
      targetLoop' fuel' passed (discards + 1) seeds
      end
      end in
      targetLoop' fuel 0 0 seeds pool utility.
    \end{rocqcodesmaller}
    \captionof{figure}{Custom-Feedback Guided (Targeted) Testing Loop in Rocq}
    \label{fig:targetloop}
  \end{minipage}
  \begin{minipage}{0.39\textwidth}
    The structure of the targeted property runner depicted in
    Fig.~\ref{fig:targeted-property-runner} is also reflected in the
    structure of the {\tt targetLoop} in Fig.~\ref{fig:targetloop}.
    This loop performs all the usual bookkeeping we discussed in the
    simple loop of Fig.~\ref{fig:simple-generational-testing-loop-rocq}, but adds mutation and
    feedback mechanisms guiding the search towards interesting
    inputs. The loop is parameterized by the \textcolor{feedback}{\tt
      feedback function}, \textcolor{seedpool}{\tt the seed pool},
    and \textcolor{seedpool}{\tt the utility function}, where the seed pool
    can be configured with different data structures such as a priority,
    FIFO, or FILO queue, and the utility function can be configured with
    different strategies such as a threshold, or a more complex
    stateful model. The loop uses the \CC{Seed Pool} and \CC{Utility}
    typeclasses to orchestrate the search.

    \vspace{2mm}

    Walking through the runner, we see that it diverges from the simple
    generational property runner in Fig.~\ref{fig:simple-generational-testing-loop-rocq}
    in its input generation, where it might
    either \overlay{generator}{generate} from scratch,
    or \overlay{mutator}{mutate} by \overlay{seedpool}{sampling} the
    seedpool. This input is then \overlay{checker}{run} through the
    property, and the failure and discarded cases are handled exactly
    the same as the simple generational property runner. If the test succeeds,
    depending on the feedback calculated by the
    user-provided feedback function, the seed might
    be \overlay{seedpool}{invested} in the seed pool, or the seed pool
    might be \overlay{seedpool}{revised} to reduce the energy of the
    seed. The loop then continues with the updated seed pool, and the
    passed and discarded counts.
  \end{minipage}

\end{figure}

\begin{figure}
  \begin{racketcodesmaller}
    (define (parallel-run-loop tests prop [num-workers (processor-count)])
    ; atomic counter for the test number
    (define counter (box 0))
    ; flag set if a thread finds a counterexample
    (define found-counterexample? (box #f))
    ; function called by each thread
    (define (worker-thunk)
    ; each thread creates its own random number generator
    (define rng (make-pseudo-random-generator))
    (let worker-loop ([passed 0]
      [discards 0])
    ; fetch and increment the test number counter
    (define n (box-faa! counter 1))
    (cond
    ; if the number of tests has exceeded the total, return the thread results
      [(>= n tests) (results #f passed discards #f)]
    ; if another thread has found a counterexample, return the thread results
      [(unbox found-counterexample?) (results #f passed discards #f)]
      [else
        ; run a single test
        (let-values ([(res env) (gen-and-run prop run-rackcheck-generator rng n)])
        (case res
        ; if a counterexample was found, set the found flag
        ; and return the thread results
          [(fail)
            (set-box! found-counterexample? #t)
            (results #t passed discards env)]
        ; on pass or discard, increment the relevant counter and recur
          [(pass) (worker-loop (add1 passed) discards)]
          [(discard) (worker-loop passed (add1 discards))]))])))
    ; spawn workers
    (define workers
    (for/list ([_ (in-range num-workers)])
    (future worker-thunk)))
    ; read results from workers
    (for/fold ([res (results #f 0 0 #f)])
    ([worker workers])
    ; get results from this worker
    (define worker-res (touch worker))
    ; combine with previous worker results
    (results (or (results-foundbug? res) (results-foundbug? worker-res))
    (+ (results-passed res) (results-passed worker-res))
    (+ (results-discards res) (results-discards worker-res))
    (or (results-counterexample res) (results-counterexample worker-res)))))
  \end{racketcodesmaller}
  \caption{Parallel Property-Based Testing Loop in Racket}
  \label{fig:par-loop-appendix}
\end{figure}

\end{document}

%% file: figures/interaction/tikz-loops.tex
\tikzset{
  component/.style={
    minimum size=0.9cm,
    draw,
    line width=1.5pt,
    font=\Large\sffamily\bfseries,
    inner sep=0pt
  },
  generator box/.style={component, draw=generator, text=generator},
  checker box/.style={component, draw=checker, text=checker},
  shrinker box/.style={component, draw=shrinker, text=shrinker},
  printer box/.style={component, draw=printer, text=printer, line width=1pt},
  mutator box/.style={component, draw=mutator, text=mutator},
  feedback box/.style={component, draw=feedback, text=feedback},
  combfeedback box/.style={component, draw=feedback, text=feedback, minimum size=0.7cm},
  size box/.style={component, draw=shrinker, text=shrinker, fill=shrinker!15, minimum size=0.8cm, font=\normalsize\bfseries},
  loop box/.style={
    draw=gray!70,
    dashed,
    line width=1pt,
    rounded corners=6pt,
    inner sep=8pt
  },
  loop region/.style={
    draw=gray!70,
    dashed,
    line width=1pt,
    rounded corners=6pt,
    fill=gray!8,
    inner sep=8pt
  },
  seed pool/.style={
    draw=seedpool,
    dashed,
    line width=1pt,
    rounded corners=4pt,
    inner sep=5pt
  },
  loop arrow/.style={
    ->,
    >={Stealth[length=2.5mm]},
    line width=0.7pt
  },
  arrow label/.style={
    font=\tiny\sffamily,
    inner sep=1pt
  },
  loop title/.style={
    font=\scriptsize\sffamily,
    text=gray!70
  }
}



%% file: macros.tex
\usepackage[utf8]{inputenc}
\usepackage[english]{babel}

\usepackage{xcolor}
\definecolor{dkgreen}{rgb}{0,0.6,0}
\definecolor{ltblue}{rgb}{0,0.4,0.4}
\definecolor{dkviolet}{rgb}{0.3,0,0.5}

\usepackage{listings}

\lstdefinelanguage{Coq}{ 
    mathescape=true,
    texcl=false, 
    morekeywords=[1]{Section, Module, End, Require, Import, Export,
        Variable, Variables, Parameter, Parameters, Axiom, Hypothesis,
        Hypotheses, Notation, Local, Tactic, Reserved, Scope, Open, Close,
        Bind, Delimit, Definition, Let, Ltac, Fixpoint, CoFixpoint, Add,
        Morphism, Relation, Implicit, Arguments, Unset, Contextual,
        Strict, Prenex, Implicits, Inductive, CoInductive, Record,
        Structure, Canonical, Coercion, Context, Class, Global, Instance,
        Program, Infix, Theorem, Lemma, Corollary, Proposition, Fact,
        Remark, Example, Proof, Goal, Save, Qed, Defined, Hint, Resolve,
        Rewrite, View, Search, Show, Print, Printing, All, Eval, Check,
        Projections, inside, outside, Def},
    morekeywords=[2]{forall, exists, exists2, fun, fix, cofix, struct,
        match, with, end, as, in, return, let, if, is, then, else, for, of,
        nosimpl, when},
    morekeywords=[3]{Type, Prop, Set, true, false, option},
    morekeywords=[4]{pose, set, move, case, elim, apply, clear, hnf,
        intro, intros, generalize, rename, pattern, after, destruct,
        induction, using, refine, inversion, injection, rewrite, congr,
        unlock, compute, ring, field, fourier, replace, fold, unfold,
        change, cutrewrite, simpl, have, suff, wlog, suffices, without,
        loss, nat_norm, assert, cut, trivial, revert, bool_congr, nat_congr,
        symmetry, transitivity, auto, split, left, right, autorewrite},
    morekeywords=[5]{by, done, exact, reflexivity, tauto, romega, omega,
        assumption, solve, contradiction, discriminate},
    morekeywords=[6]{do, last, first, try, idtac, repeat},
    morecomment=[s]{(*}{*)},
    showstringspaces=false,
    morestring=[b]",
    morestring=[d]’,
    tabsize=3,
    extendedchars=false,
    sensitive=true,
    breaklines=false,
    basicstyle=\small,
    captionpos=b,
    columns=[l]flexible,
    identifierstyle={\ttfamily\color{black}},
    keywordstyle=[1]{\ttfamily\color{dkviolet}},
    keywordstyle=[2]{\ttfamily\color{dkgreen}},
    keywordstyle=[3]{\ttfamily\color{ltblue}},
    keywordstyle=[4]{\ttfamily\color{dkblue}},
    keywordstyle=[5]{\ttfamily\color{dkred}},
    stringstyle=\ttfamily,
    commentstyle={\ttfamily\color{dkgreen}},
    literate=
    {\[\[}{$\llbracket$}1
    {\]\]}{$\rrbracket$}1
    {\\emp}{$\emptyset$}1
    {\\.}{$\cdot$}1
    {\\forall}{{\color{dkgreen}{$\forall\;$}}}1
    {\\exists}{{$\exists\;$}}1
    {<-}{{$\leftarrow\;$}}1
    {->}{{$\rightarrow\;$}}1
    {<->}{{$\leftrightarrow\;$}}1
    {<==}{{$\leq\;$}}1
    {\#}{{$^\star$}}1 
    {\\o}{{$\circ\;$}}1 
    {\@}{{$\cdot$}}1 
    {\/\\}{{$\wedge\;$}}1
    {\\\/}{{$\vee\;$}}1
    {~}{{$\sim$}}1
    {\@\@}{{$@$}}1
    {\\mapsto}{{$\mapsto\;$}}1
    {\\hline}{{\rule{\linewidth}{0.5pt}}}1
}[keywords,comments,strings]
\lstnewenvironment{rocqcode}{\lstset{language=Coq,mathescape=true,basicstyle=\ttfamily\small}}{}
\lstnewenvironment{rocqcodesmaller}{\lstset{language=Coq,mathescape=true,basicstyle=\ttfamily\footnotesize}}{}
\newcommand{\CC}{\lstinline[language=Coq, basicstyle=\ttfamily\small]}

\lstdefinestyle{haskellstyle}{
	language=Haskell,
	mathescape=true,
	basicstyle=\ttfamily\small,
	showstringspaces=false,
	columns=[l]flexible,
	identifierstyle={\ttfamily\color{black}},
	keywordstyle={\ttfamily\color{dkviolet}},
	keywordstyle=[2]{\ttfamily\color{dkblue}},
	keywordstyle=[3]{\ttfamily\color{ltblue}},
	stringstyle={\ttfamily\color{dkred}},
	commentstyle={\ttfamily\upshape\color{dkgreen}},
	morekeywords=[2]{case,of,let,in,where,if,then,else,do},
	morekeywords=[3]{IO,Int,Integer,Bool,Char,Maybe,Either},
    deletekeywords={String, StdGen, newStdGen},
}
\lstnewenvironment{hask}{\lstset{style=haskellstyle}}{}
\newcommand{\HC}{\lstinline[style=haskellstyle, basicstyle=\ttfamily\small]}

\lstdefinelanguage{racket} {
	morecomment=[l]{;},         
	morecomment=[s]{\#|}{|\#},  
	commentstyle={\color{dkgreen}\upshape\ttfamily},
	morestring=[b]",
	stringstyle=\color{red},
	literate=%
		{lambda}{{$\lambda$}}1
		{->}{{$\rightarrow$}}1,
	classoffset=1,
		morekeywords={
			define, define-syntax, define-macro, define-datatype, lambda, define-stream, stream-lambda
		},
		keywordstyle=\color{purple},
	classoffset=2,
		morekeywords={
			begin, call-with-current-continuation, call/cc, call-with-input-file, call-with-output-file, 
			cases, cond, do, else, for-each, if,
			let*, let, let-syntax, letrec, letrec-syntax,
			define-context, define-controller, Integer, Boolean, get, when-required, when-provided,
			maybe_publish, require, submod, or/c, ->, \#\%module-begin, always_publish, with-syntax, define-struct/contract, syntax-case, define/contract,
			let-values, let*-values,
			module, provide,
			and, or, not, 
			delay, force,
			\#`, \#',
			null, cons, car, cdr, cadr, caddr,
			\#lang, implement, begin-for-syntax, rename-out,
			quasiquote, quote, unquote, unquote-splicing,
			map, filter, foldl, syntax, syntax-rules, eval, environment, query,
                        match, match*, \#f, \#t, equal,
			struct
		},
		keywordstyle=\color{blue},
	classoffset=3,
		morekeywords={forall, implies, property},
		keywordstyle=\color{purple},
	classoffset=0,
	alsoletter={',`,-,/,>,<,\#,\%},
	moredelim=**[is][\color{lightgray}]{<<@<<}{>>@>>},
	moredelim=**[is][\itshape\color{OliveGreen}]{<<;<<}{>>;>>},
    columns=[l]flexible,
}

\lstnewenvironment{racketcode}{\lstset{language=racket,mathescape=true,basicstyle=\ttfamily\small}}{}
\lstnewenvironment{racketcodesmaller}{\lstset{language=racket,mathescape=true,basicstyle=\ttfamily\footnotesize}}{}
\newcommand{\RC}{\lstinline[language=racket, basicstyle=\ttfamily\small]}

%% file: figures/interaction/quickchick.tex
\begin{tikzpicture}[node distance=0.8cm]
  \node[generator box] (G) {G};
  \node[checker box, right=1.5cm of G] (C1) {C};

  \draw[loop arrow] ([yshift=3pt]G.east) to[] node[arrow label, above] {\texttt{check e}} ([yshift=3pt]C1.west);
  \draw[loop arrow] ([yshift=-3pt]C1.west) to[] node[arrow label, below] {\texttt{gen}} ([yshift=-3pt]G.east);

  \begin{scope}[on background layer]
    \node[loop region, fit=(G)(C1), inner ysep=16pt, inner xsep=6pt, label={[loop title]above:Generation Loop}] (genloop) {};
  \end{scope}

  \node[shrinker box, right=1.5cm of C1] (S) {S};
  \node[checker box, right=1.5cm of S] (C2) {C};

  \draw[loop arrow] ([yshift=3pt]S.east) to[] node[arrow label, above] {\texttt{check e'}} ([yshift=3pt]C2.west);
  \draw[loop arrow] ([yshift=-3pt]C2.west) to[] node[arrow label, below] {\texttt{shrink e | e'}} ([yshift=-3pt]S.east);

  \begin{scope}[on background layer]
    \node[loop region, fit=(S)(C2), inner ysep=16pt, inner xsep=6pt, label={[loop title]above:Shrinking Loop}] (shrinkloop) {};
  \end{scope}

  \draw[loop arrow] (C1.east) -- node[arrow label, above] {\texttt{shrink e}} (S.west);

  \node[printer box, right=1.5cm of C2] (P) {P};

  \draw[loop arrow] (C2.east) -- node[arrow label, above] {\texttt{print e/e'}} (P.west);
\end{tikzpicture}

%% file: figures/interaction/fuzzing.tex
\begin{tikzpicture}[node distance=0.8cm]
  \node[generator box] (G) {G};
  \node[mutator box, below=0.4cm of G] (M) {M};

  \begin{scope}
    \node[seed pool, fit=(G)(M), inner sep=3pt, label={[font=\scriptsize, text=seedpool]above:Seed Pool}] (seedpool) {};
  \end{scope}

  \node[checker box, right=1.5cm of seedpool] (C1) {C};

  \node[feedback box, fit=(C1), inner sep=1pt] (F) {};

  \draw[loop arrow] ([yshift=3pt]seedpool.east) to[] node[arrow label, above] {\texttt{check e}} ([yshift=3pt]C1.west);
  \draw[loop arrow] ([yshift=-3pt]C1.west) to[] node[arrow label, below] {\texttt{gen/mutate f}} ([yshift=-3pt]seedpool.east);

  \begin{scope}[on background layer]
    \node[loop region, fit=(seedpool)(F), inner ysep=14pt, inner xsep=4pt, label={[loop title]above:Fuzzing Loop}] (fuzzloop) {};
  \end{scope}

  \node[shrinker box, right=1.3cm of C1] (S) {S};
  \node[checker box, right=1.5cm of S] (C2) {C};

  \draw[loop arrow] ([yshift=3pt]S.east) to[] node[arrow label, above] {\texttt{check e'}} ([yshift=3pt]C2.west);
  \draw[loop arrow] ([yshift=-3pt]C2.west) to[] node[arrow label, below] {\texttt{shrink e' | e}} ([yshift=-3pt]S.east);

  \begin{scope}[on background layer]
    \node[loop region, fit=(S)(C2), inner ysep=16pt, inner xsep=6pt, label={[loop title]above:Shrinking Loop}] (shrinkloop) {};
  \end{scope}

  \draw[loop arrow] (C1.east) -- node[arrow label, above] {\texttt{shrink e}} (S.west);

  \node[printer box, right=1.2cm of C2] (P) {P};

  \draw[loop arrow] (C2.east) -- node[arrow label, above] {\texttt{print e/e'}} (P.west);
\end{tikzpicture}

%% file: figures/interaction/rackcheck.tex
\begin{tikzpicture}[node distance=0.8cm]
  \node[generator box] (G1) {G};
  \node[checker box, right=1.2cm of G1] (C1) {C};

  \draw[loop arrow] ([yshift=3pt]G1.east) to[] node[arrow label, above] {\texttt{check e}} ([yshift=3pt]C1.west);
  \draw[loop arrow] ([yshift=-3pt]C1.west) to[] node[arrow label, below] {\texttt{gen\textsubscript{bytes}}} ([yshift=-3pt]G1.east);

  \begin{scope}[on background layer]
    \node[loop region, fit=(G1)(C1), inner ysep=16pt, inner xsep=6pt, label={[loop title]above:Generation Loop}] (genloop) {};
  \end{scope}

  \node[generator box, right=2cm of C1] (G2) {G};
  \node[checker box, right=1.2cm of G2] (C2) {C};

  \draw[loop arrow] ([yshift=3pt]G2.east) to[] node[arrow label, above] {\texttt{check e'}} ([yshift=3pt]C2.west);
  \draw[loop arrow] ([yshift=-3pt]C2.west) to[] node[arrow label, below] {\texttt{gen\textsubscript{bytes'}}} ([yshift=-3pt]G2.east);

  \begin{scope}[on background layer]
    \node[loop region, fit=(G2)(C2), inner ysep=16pt, inner xsep=6pt, label={[loop title]above:Integrated Shrinking Loop}] (shrinkloop) {};
  \end{scope}

  \draw[loop arrow] (C1.east) -- node[arrow label, above] {\texttt{gen\textsubscript{bytes}}} (G2.west);

  \node[printer box, right=1.5cm of C2] (P) {P};

  \draw[loop arrow] (C2.east) -- node[arrow label, above] {\texttt{print e/e'}} (P.west);
\end{tikzpicture}

%% file: figures/interaction/targeted.tex
\begin{tikzpicture}[node distance=0.8cm]
  \node[generator box] (G) {G};
  \node[mutator box, below=0.4cm of G] (M) {M};

  \begin{scope}
    \node[seed pool, fit=(G)(M), inner sep=3pt, label={[font=\scriptsize, text=seedpool]above:Seed Pool}] (seedpool) {};
  \end{scope}

  \node[feedback box, right=1.5cm of seedpool, yshift=-3mm] (F) {F};

  \node[checker box, right=1.5cm of F, , yshift=3mm] (C1) {C};

  \draw[loop arrow] ([yshift=9pt]seedpool.east) to[] node[arrow label, above] {\texttt{check e}} ([yshift=9pt]C1.west);
  \draw[loop arrow] ([yshift=-9pt]C1.west) to[] node[arrow label, below] {\texttt{feedback e}} ([yshift=0pt]F.east);
  \draw[loop arrow] ([yshift=0pt]F.west) to[] node[arrow label, below] {\texttt{gen/mutate f}} ([yshift=-9pt]seedpool.east);

  \begin{scope}[on background layer]
    \node[loop region, fit=(seedpool)(F)(C1), inner ysep=14pt, inner xsep=4pt, label={[loop title]above:Targeting Loop}] (fuzzloop) {};
  \end{scope}

  \node[shrinker box, right=1.3cm of C1] (S) {S};
  \node[checker box, right=1.5cm of S] (C2) {C};

  \draw[loop arrow] ([yshift=3pt]S.east) to[] node[arrow label, above] {\texttt{check e'}} ([yshift=3pt]C2.west);
  \draw[loop arrow] ([yshift=-3pt]C2.west) to[] node[arrow label, below] {\texttt{shrink e' | e}} ([yshift=-3pt]S.east);

  \begin{scope}[on background layer]
    \node[loop region, fit=(S)(C2), inner ysep=14pt, inner xsep=6pt, label={[loop title]above:Shrinking Loop}] (shrinkloop) {};
  \end{scope}

  \draw[loop arrow] (C1.east) -- node[arrow label, above] {\texttt{shrink e}} (S.west);

  \node[printer box, right=1.2cm of C2] (P) {P};

  \draw[loop arrow] (C2.east) -- node[arrow label, above] {\texttt{print e/e'}} (P.west);
\end{tikzpicture}

%% file: figures/interaction/combinatorial.tex
\begin{tikzpicture}[node distance=0.8cm]
  \node[generator box] (G) {G};
  \node[checker box, right=1.5cm of G] (C1) {C};

  \node[combfeedback box, below=2cm of G] (CF) {CF};

  \draw[loop arrow] (C1.west) to[] node[arrow label, above] {\texttt{gen}} (G.east);

  \draw[loop arrow] ([xshift=-10pt]G.south) -- node[font=\scriptsize] {} ([xshift=-10pt]CF.north);
  \draw[loop arrow] ([xshift=-5pt]G.south) -- node[font=\scriptsize] {} ([xshift=-5pt]CF.north);
  \draw[loop arrow] ([xshift=-0pt]G.south) -- node[font=\scriptsize] {} ([xshift=-0pt]CF.north);
  \draw[loop arrow] ([xshift=5pt]G.south) -- node[font=\scriptsize] {} ([xshift=5pt]CF.north);
  \draw[loop arrow] ([xshift=10pt]G.south) -- node[arrow label, right, font=\scriptsize] {\texttt{pick e1..e5}} ([xshift=10pt]CF.north);
  
  \draw[loop arrow] 
      (CF.east)
      -| node[arrow label, xshift=-6mm, yshift=2mm, font=\scriptsize] {\texttt{check e}} (C1.south);

  \begin{scope}[on background layer]
    \node[loop region, fit=(G)(C1)(CF), inner ysep=10pt, inner xsep=6pt, label={[loop title]above:Generation Loop}] (genloop) {};
  \end{scope}

  \node[shrinker box, right=1.5cm of C1] (S) {S};
  \node[checker box, right=1.5cm of S] (C2) {C};

  \draw[loop arrow] ([yshift=3pt]S.east) to[] node[arrow label, above] {\texttt{check e'}} ([yshift=3pt]C2.west);
  \draw[loop arrow] ([yshift=-3pt]C2.west) to[] node[arrow label, below] {\texttt{shrink e | e'}} ([yshift=-3pt]S.east);

  \begin{scope}[on background layer]
    \node[loop region, fit=(S)(C2), inner ysep=10pt, inner xsep=6pt, label={[loop title]above:Shrinking Loop}] (shrinkloop) {};
  \end{scope}

  \draw[loop arrow] (C1.east) -- node[arrow label, above] {\texttt{shrink e}} (S.west);

  \node[printer box, right=1.5cm of C2] (P) {P};

  \draw[loop arrow] (C2.east) -- node[arrow label, above] {\texttt{print e/e'}} (P.west);
\end{tikzpicture}

%% file: figures/interaction/parallel.tex
\begin{tikzpicture}[node distance=0.6cm]
  \node[size box] (Size) {Size};

  \foreach \i/\yoff in {1/1.5, 2/0.0, 3/-1.5} {
    
    \node[generator box, minimum size=0.7cm, font=\normalsize\bfseries, right=2.2cm of Size, yshift=\yoff cm] (G\i) {G};
    \node[checker box, minimum size=0.7cm, font=\normalsize\bfseries, right=1.5cm of G\i] (C\i) {C};
    
    \node[loop box, fit=(G\i)(C\i), inner sep=6pt] (GCpair\i) {};

    \draw[loop arrow, line width=0.5pt] ([yshift=2pt]G\i.east) to[] node[arrow label, above, font=\tiny] {\texttt{check e}} ([yshift=2pt]C\i.west);
    \draw[loop arrow, line width=0.5pt] ([yshift=-2pt]C\i.west) to[] node[arrow label, below, font=\tiny] {\texttt{gen}} ([yshift=-2pt]G\i.east);
  }

  \draw[loop arrow, line width=0.5pt] ([xshift=3pt]Size.north) to[bend left=20] node[arrow label, xshift=0.5cm, font=\tiny, pos=0.3] {size} ([yshift=-3pt]GCpair1.west);
  \draw[loop arrow, line width=0.5pt] ([yshift=3pt]GCpair1.west) to[bend right=20] node[arrow label, xshift=-1cm, font=\tiny, pos=0.3] {increase size} ([xshift=-3pt]Size.north);
  \draw[loop arrow, line width=0.5pt] ([yshift=-3pt]Size.east) to[] node[arrow label, yshift=-2mm, font=\tiny] {size} ([yshift=-3pt]GCpair2.west);
  \draw[loop arrow, line width=0.5pt] ([yshift=3pt]GCpair2.west) to[] node[arrow label, yshift=2mm, font=\tiny] {increase size} ([yshift=3pt]Size.east);
  \draw[loop arrow, line width=0.5pt] ([xshift=3pt]Size.south) to[bend right=20] node[arrow label, xshift=0.5cm, font=\tiny, pos=0.3] {size} ([yshift=3pt]GCpair3.west);
  \draw[loop arrow, line width=0.5pt] ([yshift=-3pt]GCpair3.west) to[bend left=20] node[arrow label, xshift=-1cm, font=\tiny, pos=0.3] {increase size} ([xshift=-3pt]Size.south);

  \begin{scope}[on background layer]
    \node[loop region, fit=(Size)(G1)(C1)(G3)(C3), inner ysep=12pt, inner xsep=8pt, label={[loop title]above:Parallel Testing Loop}] (parloop) {};
  \end{scope}

  \node[shrinker box, right=2cm of C2] (S) {S};
  \node[checker box, right=1.5cm of S] (C5) {C};

  \draw[loop arrow] ([yshift=3pt]S.east) to[] node[arrow label, above, font=\scriptsize] {\texttt{check e'}} ([yshift=3pt]C5.west);
  \draw[loop arrow] ([yshift=-3pt]C5.west) to[] node[arrow label, below, font=\scriptsize] {\texttt{shrink e | e'}} ([yshift=-3pt]S.east);

  \begin{scope}[on background layer]
    \node[loop region, fit=(S)(C5), inner ysep=14pt, inner xsep=8pt, label={[loop title]above:Shrinking Loop}] (shrinkloop) {};
  \end{scope}

  \draw[loop arrow] (parloop.east) -- node[arrow label, above, font=\scriptsize] {\texttt{shrink e}} (S.west);

  \node[printer box, right=1.5cm of C5] (P) {P};

  \draw[loop arrow] (C5.east) -- node[arrow label, above, font=\scriptsize] {\texttt{print e/e'}} (P.west);
\end{tikzpicture}